\documentclass[pra,aps,twocolumn,showpacs,superscriptaddress,nofootinbib,floatfix]{revtex4-1}
\usepackage[margin=1in]{geometry}
\usepackage[utf8]{inputenc}
\usepackage[T1]{fontenc}
\usepackage{graphicx}
\usepackage{physics}
\usepackage{amsfonts}
\usepackage{algorithm}
\usepackage{algpseudocode}
\usepackage{xcolor}
\newcommand{\Graph}{{\mathcal{G}}}
\newcommand{\Sum}{\displaystyle \sum}
\newcommand{\edges}{{\mathcal{E}}}
\newcommand{\vertices}{{\mathcal{V}}}
\newcommand{\Ham}{\hat{\mathcal{H}}}
\usepackage{multirow}
\newcommand{\alphab}{\boldsymbol{\alpha}}

\usepackage{hhline}
\usepackage[colorlinks=true, breaklinks=true, linkcolor=blue, citecolor=blue, urlcolor=blue]{hyperref}

\begin{document}
\title{
Quantum Feature Maps for Graph Machine Learning on a Neutral Atom Quantum Processor
}
\author{Boris Albrecht}
\thanks{These authors contributed equally to this work}
\affiliation{PASQAL, 7 rue Léonard de Vinci, 91300 Massy, France}
\author{Constantin Dalyac}
\thanks{These authors contributed equally to this work}
\affiliation{PASQAL, 7 rue Léonard de Vinci, 91300 Massy, France}
\affiliation{LIP6, CNRS, Sorbonne Université, 4 Place Jussieu, 75005 Paris, France}
\author{Lucas Leclerc}
\thanks{These authors contributed equally to this work}
\affiliation{PASQAL, 7 rue Léonard de Vinci, 91300 Massy, France}
\affiliation{Universit\'e Paris-Saclay, Institut d'Optique Graduate School,CNRS, Laboratoire Charles Fabry, 91127 Palaiseau, France}
\author{Luis Ortiz-Gutiérrez}
\thanks{These authors contributed equally to this work}
\affiliation{PASQAL, 7 rue Léonard de Vinci, 91300 Massy, France}
\author{Slimane Thabet}
\thanks{These authors contributed equally to this work}
\affiliation{PASQAL, 7 rue Léonard de Vinci, 91300 Massy, France}
\affiliation{LIP6, CNRS, Sorbonne Université, 4 Place Jussieu, 75005 Paris, France}
\author{Mauro D'Arcangelo}
\author{Vincent E. Elfving}
\author{Lucas Lassablière}
\author{Henrique Silvério}
\author{Bruno Ximenez}
\author{Louis-Paul Henry}
\author{Adrien Signoles}
\author{Loïc Henriet}
\email{loic@pasqal.com}
\affiliation{PASQAL, 7 rue Léonard de Vinci, 91300 Massy, France}

\date{\today}
\begin{abstract}
Using a quantum processor to embed and process classical data enables the generation of correlations between variables that are inefficient to represent through classical computation. A fundamental question is whether these correlations could be harnessed to enhance learning performances on real datasets. Here, we report the use of a neutral atom quantum processor comprising up to $32$ qubits to implement machine learning tasks on graph-structured data. To that end, we introduce a quantum feature map to encode the information about graphs in the parameters of a tunable Hamiltonian acting on an array of qubits. Using this tool, we first show that interactions in the quantum system can be used to distinguish non-isomorphic graphs that are locally equivalent. We then realize a toxicity screening experiment, consisting of a binary classification protocol on a biochemistry dataset comprising $286$ molecules of sizes ranging from $2$ to $32$ nodes, and obtain results which are comparable to those using the best classical kernels. Using techniques to compare the geometry of the feature spaces associated with kernel methods, we then show evidence that the quantum feature map perceives data in an original way, which is hard to replicate using classical kernels. 
\end{abstract}

\maketitle

\section*{Introduction}

Representing data in the form of graphs is ubiquitous in many domains of sciences. They naturally describe relationships in social networks\,\cite{Freeman2000VisualizingSN}, characterize interactions of proteins and genes\,\cite{Theocharidis2009NetworkVA} and can represent the structure of sentences in linguistics\,\cite{sole2001small}. Many impactful applications arise from efficient graph-based methods, such as predicting potential edges in recommendation systems\,\cite{Schafer01}, detecting frauds in communication networks\,\cite{Pourhabibi20}, or for protein function prediction\,\cite{Muzio20}.

While graphs offer a rich structure for manipulating complex data, the level of freedom they afford can lead to resource-consuming data analyses. It is therefore essential to create efficient machine learning (ML) models that correctly and effectively learn and extract information from graph structures. One thus often resorts to graph embedding techniques\,\cite{goyal2018graph}, which refer to finding a representation of a graph or of its individual nodes in a vector space. By finding node representatives which preserve different types of relational information from the graph, node embedding can be used for prediction tasks at the node-level, such as node classification\,\cite{Bhagat2011} or link prediction\,\cite{Liben-Nowell03}. Graph-level embeddings can be used to distinguish graphs of different nature. Notions of distances and similarities between the representative vectors can then be used to find the best boundary between datapoints with different labels in the context of supervised machine learning.  \\

Using the exponentially large Hilbert space accessible to a quantum computer in order to generate graph embeddings is an appealing idea, with many proposals and theoretical studies over the past few years\,\cite{Schuld19,Havlek19,schuld2020measuring,kishi21}. With the recent advances in geometric quantum machine learning, works have shown how graph-structured data could be encoded into quantum states and manipulated for classification, clustering or regression tasks. These efforts started with quantum convolutional neural networks \cite{Cong2019, Zheng2021}, and attempts were made to translate classical Graph Neural Network (GNN) architectures to Quantum Neural Networks (QNN) \cite{Verdon2019}. Some of the authors of the current work introduced the Quantum Evolution Kernel (QEK) approach in 2021 \cite{Henry21}, which is based on evolving a quantum register over alternating layers of (graph-encoding) Hamiltonians, and training for classification tasks. Follow-up work from the community offered generalizations of this paradigm\,\cite{EQGC2022}. Theoretical studies of geometrical quantum machine learning and their invariant properties include \cite{Larocca2022,Skolik2022}, the latter studying applications to weighted graphs. More recently, in-depth theoretical studies of equivariant and geometric quantum machine learning aspects were presented \cite{Alamos1,Alamos2,Alamos3}. Here, we specifically focus on a QEK-type quantum feature map \cite{Henry21} for graph-structured data that we experimentally investigate for various learning tasks on a $32$-qubit neutral atom Quantum Processing Unit (QPU).\\

The structure of the paper is as follows: we first introduce the concept of a graph quantum feature map using neutral atom technology in Sec.\,\ref{sec:embeddings}. We then assess in Sec.\,\ref{sec:expr_feature_map} its expressive power by showing that it enables to distinguish two graphs that are locally equivalent but non-isomorphic\,\cite{WL68}.  In Sec.\,\ref{sec:Binary_class}, we experimentally realize a quantum graph kernel on a real-world classification task to predict toxicity for a dataset of molecules (PTC-FM)\,\cite{Helma2001}. Importantly, we benchmark our approach by comparing its performance with several classical kernels in Sec.\,\ref{sec:classification_results}. Finally, we evaluate the potential advantage of our method in Sec.\,\ref{sec:geo_diff} by means of a novel metric that is sensitive to the similarity between the geometry of the feature spaces of two kernels\,\cite{Huang2021}. This is a strong indication of the potential of the method and of its capacity to capture new features that classical kernels would miss.

\section{Quantum feature map for graph-structured data}
\label{sec:embeddings}
\begin{figure}[!b]
	\centering
	\includegraphics[width=1\linewidth]{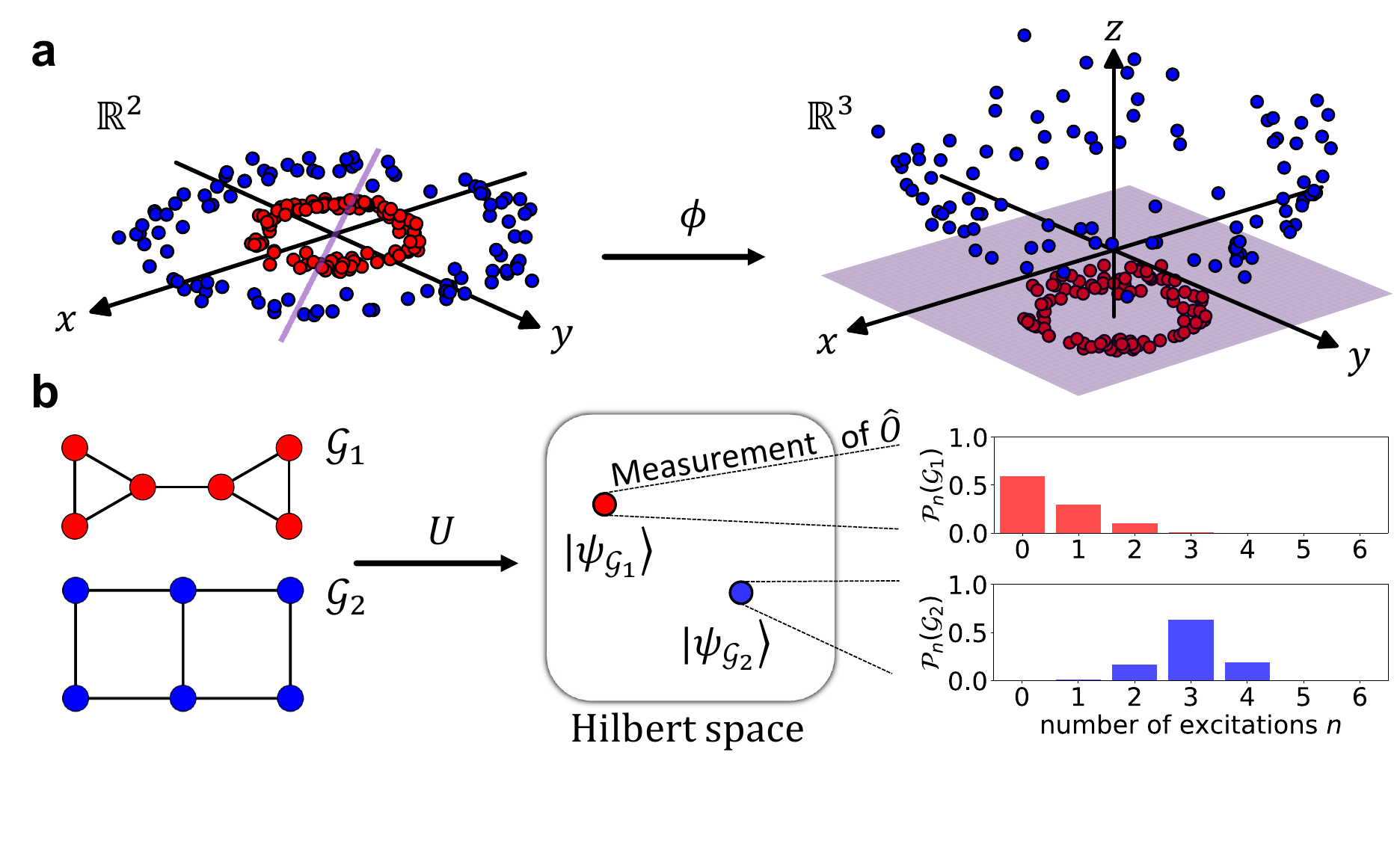}
	\caption{
	\textbf{a}. We seek to find a binary classifier enabling the separation of the two data classes by a hyperplane (purple). After a transformation $\phi((x, y)) = (x, y, x^2 + y^2)$ mapping the data point from $\mathbb{R}^2$ to  $\mathbb{R}^3$, the transformed data points become linearly separable. \textbf{b}. Illustration of the effect of a quantum feature map $U(\Graph;t)$ and subsequent measurements for graph-structured data. A parameterized quantum unitary $U$ is applied to atomic registers arranged under the form of UD graphs. Experimentally measured observable distributions are then used for learning tasks. }
	\label{fig:featuremap}
\end{figure}
In many classical machine learning methods, one seeks to map input data into a different space, making it easier to work with. An example is shown in Fig.\,\,\ref{fig:featuremap}\textbf{a}, where we illustrate how to easily solve a binary classification task for data points in a two-dimensional plane by embedding them into three-dimensional space. After a transformation $\phi$ taking data points from $\mathbb{R}^2$ to $\mathbb{R}^3$, the resulting vectors can be easily separated by an horizontal plane. The class of any new data point can then be directly deduced from its location relative to this plane. In Quantum Machine Learning\,\cite{Schuld19,Havlicek2019}, one typically uses the Hilbert space associated with a set of qubits as the embedding space. Such an embedding is usually built from the dynamics of a quantum system depending on the input data as well as external variational parameters.\\

In this paper, we use a neutral-atom based QPU made of single $^{87}$Rb atoms trapped in arrays of optical tweezers~\cite{barredo_synthetic_2018,Nogrette14,browaeys2020many,Henriet2020quantum,Morgado2021}. The qubits are encoded into the ground state $\ket{0}=\ket{5S_{1/2},F=2,m_F=2}$ and a Rydberg state $\ket{1}=\ket{60S_{1/2}, m_J=1/2}$. This effective two-level system is addressed with a two-photon laser excitation through an intermediate state $6P_{3/2}$. The first (respectively second) photon excitation is generated by a $420$-nm ($1013$-nm) laser beam. Both of them are far-detuned from their addressed transitions, so as to ensure negligible effect of the intermediate state. At the end of the laser sequence, the state of the atomic qubits is read out by fluorescence imaging (see Appendix\, \ref{app:ES}).

When promoted to Rydberg states, the atoms behave as large electric dipoles and experience strong van der Waals interactions. The dynamics of a set of $N$ qubits at positions $\{\mathbf{r}_i\}_{i=1\hdots N}$ is thus governed by the following Hamiltonian:
\begin{equation}
   \Ham  = \hbar\sum_{i=1}^{N}  \left(\frac{\Omega }{2} \hat \sigma_i^x - \delta\,\hat n_i\right) + \sum_{i<j} \frac{C_6}{|\mathbf r_i - \mathbf r_j |^6} \hat n_i \hat n_j
   \label{eq:Ham}
\end{equation}
where $\hat{\sigma}_i^\alpha$
are Pauli matrices and $\hat{n}_i=\left(1+\hat{\sigma}_i^z\right)/2$. $|\mathbf r_i - \mathbf r_j |$ is the distance between qubits $i$ and $j$, and $C_6/h \simeq 138 $ GHz$\cdot \mu {\rm m}^6$ for the Rydberg state considered. Controlling both intensities and frequencies of each laser field, we can effectively drive the qubit register uniformly with time-dependent tunable Rabi frequency $\Omega$ and detuning $\delta$.

Key to our study is the programmability of the qubit register's geometry. In neutral atom processors, one can modify the spatial arrangement of qubits\,\cite{barredo_atom-by-atom_2016,barredo_synthetic_2018} and reproduce the geometrical shape of various graphs with atoms in tweezers. We will restrict ourselves to a set of graphs called Unit Disk (UD) graphs, for which two nodes in the plane are connected by an edge if the distance between them is smaller than a given threshold. UD graphs are intimately related to Rydberg physics through the mechanism of Rydberg blockade\,\cite{Jaksch_2000,Gaetan2009,browaeys2020many}, where an atom excited to a Rydberg state prevents other neighboring atoms to be excited within a certain blockade radius. For an atomic register reproducing a UD graph $\Graph$, the $1/r^6$ power law of the van der Waals interactions effectively restricts, in a good approximation, the summation in the third term of Eq.\,\eqref{eq:Ham} to pairs of indices $(i,j)$ sharing an edge in $\Graph$. The topology of the interaction term in Eq.\,\eqref{eq:Ham} then becomes the one of the graph under consideration, giving rise to a graph-dependent Hamiltonian $\Ham_\Graph$. This property has notably been harnessed for solving combinatorial graph optimization problems\,\cite{Pichler18,Henrietrobustness,Dalyac2021,MinhThi2022,Ebadi22optim,Kim2022,Byun22,Wurtz2022,Dalyac22}.\\


Starting from a UD graph $\Graph$ reproduced in the array of tweezers with qubits all starting in $\ket{0}$, we apply a parameterized unitary transformation to generate a wavefunction $|\psi_\Graph \rangle $ of the form 
\begin{equation}
    |\psi_\Graph \rangle = U(\Graph;t)\,|0\rangle^{\otimes |\Graph|},
\end{equation}
where we define the time-evolution operator $U(\Graph;t)=\mathcal{T}\left[\exp(-i/\hbar\int_{s=0}^t \hat{\mathcal{H}}_\Graph (s)ds)\right]$ to be our \textit{quantum feature map} unitary for graph-structured data. Throughout this paper, we will restrict ourselves to laser pulses with constant detuning $\delta$ and Rabi frequency $\Omega$, with an adjustable duration $t$. Depending on the task at hand, we consider various observables $\hat O$ to evaluate on $\ket{\psi_\Graph}$. Measurements of a site-dependent (respectively global) observable give rise to a probability distribution $\mathcal{P}$ which is node (graph) specific and can be used for various machine learning tasks at the node (graph) level. In the following, we show theoretically and experimentally that the graph quantum feature map already shows interesting properties when associated with local or global observables built from single-body expectation values $\langle \hat{O}_{j=1,\hdots,\abs{\Graph}} \rangle$.

\begin{figure*}[t!]
    \centering
	\includegraphics[width=1
	\linewidth]{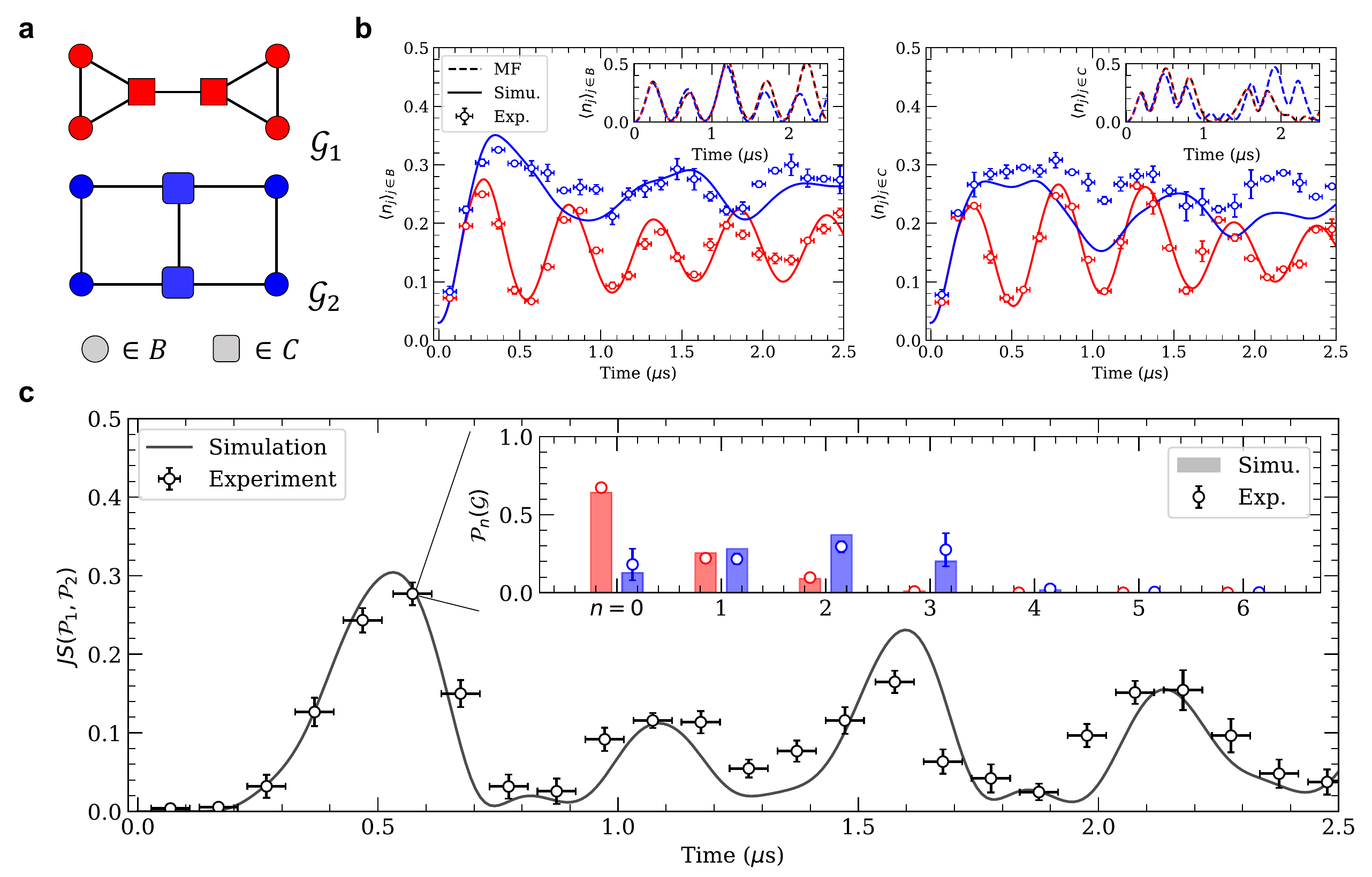}
	\caption{
	\textbf{a}. $\mathcal{G}_1$ (red) and $\mathcal{G}_2$ (blue) are two different graphs with identical local structure. Based on their neighborhood, the nodes either belong to the \textit{border} $B$  (circle) or to the \textit{center} $C$ (square). \textbf{b}. We plot the evolution of the mean occupation $\langle n_i\rangle$ of the two regions $B$ (left) and $C$ (right) for both graphs $\mathcal{G}_1$ (red) and $\mathcal{G}_2$ (blue). The dots represent the experimental results while the full curves show noisy simulation results. Horizontal error bars account for the sequence-trigger uncertainty ($\approx 40$ns) while the vertical ones account for the sampling noise. The insets show the corresponding mean field dynamics (dashed) with only NN (black) or full (colored) interactions.  \textbf{c}. The evolution of the Jensen-Shannon divergence obtained experimentally (dot) is compared to the noisy simulation (plain). At each point in time, $JS(\mathcal{P}_1,\mathcal{P}_2)$ is computed using the excitation distributions $\mathcal{P}_{1/2}=\{\mathcal{P}_n(\Graph_{1/2})\}_{n=0\hdots 6}$ obtained either numerically (bar) or experimentally (dot). The inset depicts $\mathcal{P}_{1/2}$ obtained at $t\approx0.57~\mu$s which yields the maximum value $JS_{max}\approx0.28$ reached.}
	\label{fig:res_WL_test}
\end{figure*}

\section{Expressive power of the graph quantum feature map}
\label{sec:expr_feature_map}

The graph quantum feature map already shows interesting properties when associated with single-body observables $\langle\hat{O}_{j=1,\hdots,\abs{\Graph}}\rangle$. The measured values are not only affected by local graph properties such as node degrees, but also by more global ones such as the presence of cycles. This enrichment provided by the quantum dynamics contrasts with the locality of node representations in many classical graph machine learning. This key feature comes from the fact that the quantum dynamics of a given spin model ({\it e.g.} an Ising model) will be significantly influenced, beyond short times (given by the Lieb-Robinson bound \cite{LiebRobinson, LiebRobinson_QC}), by the complete structure of the graph. 

We illustrate experimentally this behavior for two graphs $\Graph_1$ and $\Graph_2$ that are non-isomorphic but locally identical. In these graphs, nodes can be separated into two equivalence classes according to their neighborhood: border nodes $B$  have one degree-3 neighbor and one degree-2 neighbor, while center nodes $C$ have two degree-2 neighbors and one degree-3 neighbor (see Fig.\,\,\ref{fig:res_WL_test}\textbf{a}). 

We first map the graphs in a tweezers array with a nearest-neighbor (NN) distance of $r_{NN} = 5.3\,\mu$m and apply a constant pulse with $\Omega/2\pi= 1.0\,\text{MHz}$ and $\delta/2\pi=0.7\,\text{MHz}$. We then measure the local mean Rydberg excitation $\langle n_j\rangle_{j \in B/C}$ for varying pulse duration $t \in [ 0, 2.5 ]\,\mu$s. As illustrated in Fig.\,\,\ref{fig:res_WL_test}\textbf{b}, a qualitative difference in the dynamics of both graph appears after $t \sim 0.25\,\mu$s. Precisely, the excitation of the border nodes (see Fig.\,\,\ref{fig:res_WL_test}\textbf{b}, left panel) is initially increasing with indistinguishable behavior between the two graphs. Then, a distinction appears between the two graph instances. The mean density for the border qubits of $\Graph_1$ exhibits damped oscillations around $\langle n_B \rangle \sim 0.15$ with period of the order of $0.5\,\mu$s while for $\Graph_2$ it exhibits flatter oscillations centered around $0.25$ with period around $1\,\mu$s. We can observe a comparable distinction between the two graphs for the center qubits (see Fig.\,\,\ref{fig:res_WL_test}\textbf{b}, right panel). The experimental measurements are consistent with the theoretical predictions and with the expected level of noise (see Appendix\,\ref{app:noise_model} for more details). 

When restricted to the mean-field approximation (or similarly in the classical limit), the qubits' dynamics on either graphs are far more similar, as illustrated in the insets of panels in Fig.\,\,\ref{fig:res_WL_test}\-{\bf b}. We still observe distinct dynamics between the two graphs, which is due to next nearest neighbors (NNN) interactions (more pronounced for the center nodes). If we neglected those NNN interactions, the mean-field equations governing the dynamics of each qubit would only depend on its direct neighborhood, {\it i.e.} the local structure of the graph.  In that case, the qubits dynamics for $\Graph_1$ and $\Graph_2$ obey the exact same equations (see black dashed line in the insets of Fig.\,\,\ref{fig:res_WL_test}\-{\bf b}).  We therefore conclude that the presence of interactions in the system enables us to discriminate between the two non-isomorphic graphs $\Graph_1$ and $\Graph_2$ by evaluating node-level local observables $\langle n_B \rangle$ or $\langle n_C \rangle$.

By looking at $\hat{O}= \sum_{i=1}^6 \hat n_i$, we can more quantitatively quantify the difference in the dynamics between the two graphs. To this end, we first compute the histogram $\mathcal{P}_i$ of number of excitations observed in each shot on graph $\Graph_i$. The difference between those graphs is then estimated via the Jensen-Shannon divergence of their respective histograms, a commonly used distance measure between probability distributions: 
\begin{equation}
    \label{eq:JS}
	JS(\mathcal{P}_1, \mathcal{P}_2) = H\left(\frac{\mathcal{P}_1+\mathcal{P}_2}{2}\right) -\frac{H(\mathcal{P}_1)+H(\mathcal{P}_2)}{2}.
\end{equation}
Here $H(\mathcal{P})=-\sum_k p_k\log p_k$ is the Shannon entropy of $\mathcal{P}=(p_1, \dots p_{|\Graph|})$. $JS(\mathcal{P}_1, \mathcal{P}_2)$ takes values in $\left[0,\log 2\right]$, with $JS(\mathcal{P}, \mathcal{P})=0$, and $JS$ is maximal if $\mathcal{P}_1$ and $\mathcal{P}_2$ have disjoint supports.
This is illustrated in Fig.\,\,\ref{fig:res_WL_test}{\bf c}, where the largest difference $JS_{max}\approx 0.28$ is achieved (roughly 40\% of the maximal value) at a time $t\sim 0.57 \mu s$. At this duration, the distribution for $\Graph_1$ is sharply peaked at $n=0$ while that of $\Graph_2$ is wider and peaks around $n=2$, as illustrated in the inset of Fig.\,\,\ref{fig:res_WL_test}\textbf{c}. We note that the local observables $\langle n_j\rangle_{j \in B/C}$ exhibit maximal deviation at this same duration $t$, indicating direct correspondence between measurements at the node and graph levels.

The dependency of local observables evaluated after the application of the quantum feature map on global graph structures have interesting consequences regarding quantum-enhanced versions of GNN\,\cite{Verdon19,EQGC2022,Thabet2022}. Standard GNN architectures are based on the concept of Message Passing \cite{pmlr-v70-gilmer17a,WU21}, in which the information can only be propagated locally on the graphs nodes (see Appendix \ref{app:MPNN} for details). Incorporating a propagation rule built from the quantum feature map above would enables us to go beyond this well-known limitation of GNN architectures.

\begin{figure*}[t!]
	\centering
	\includegraphics[width=1\textwidth]{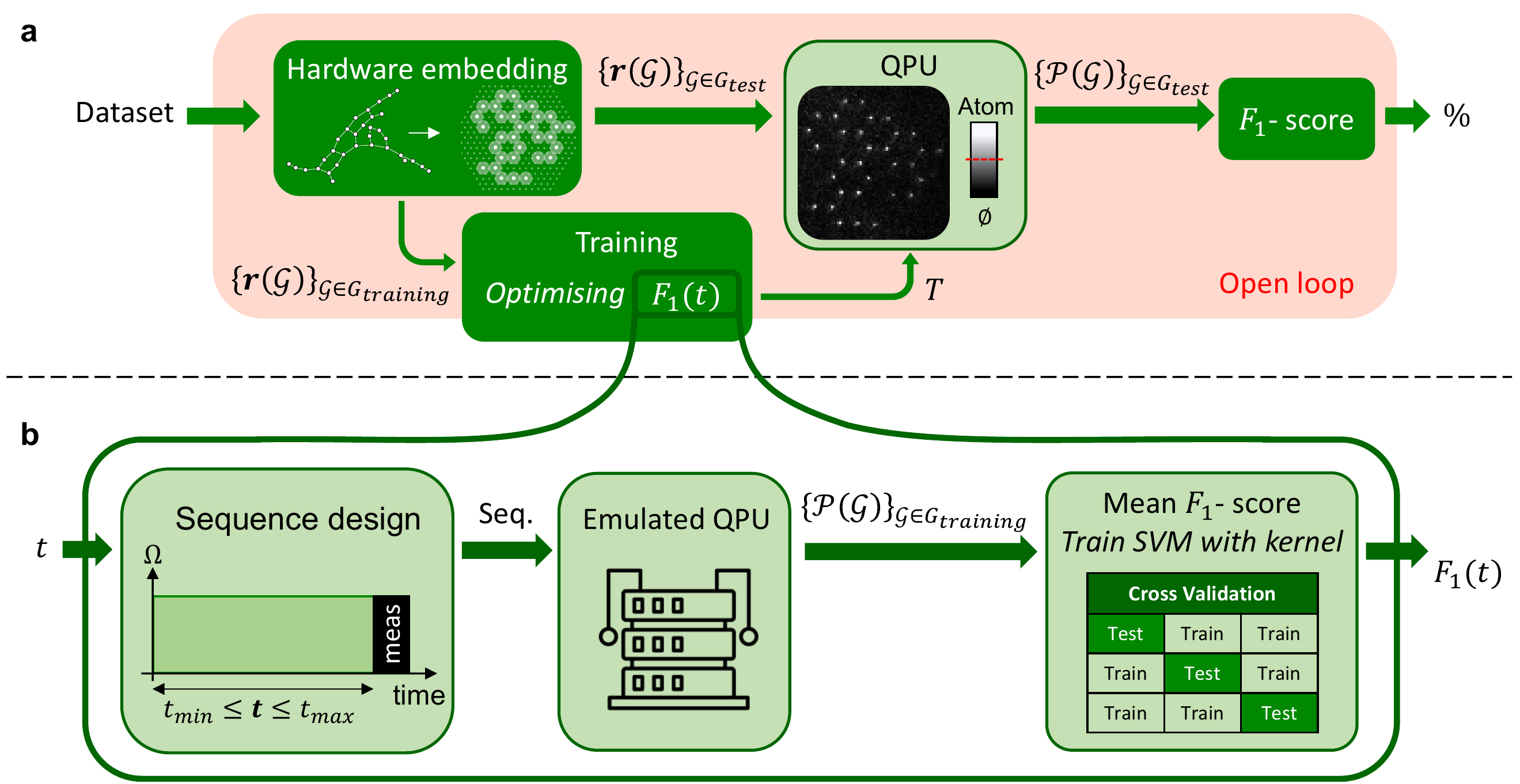}
	\caption{
	\textbf{a}. A dataset of graphs $\Graph$ is first mapped onto atomic registers $\textbf{r}(\Graph)$ implementable on the QPU, and separated between a training set $\mathcal{G}_{training}$ and a test set $\mathcal{G}_{test}$.  We use the training set to determine numerically the optimal pulse sequence to be applied on the hardware using a grid search algorithm for optimizing $F_1(t)$ (see \textbf{b}).  This training phase outputs the optimal parameter $T$ used for the laser-pulse sequence applied experimentally on each register of the test set. $F_1$ is then derived from the measured probability distributions $\{\mathcal{P}(\Graph)\}_{\Graph\in G_{test}}$. \textbf{b.} The optimization of the score function $F_1$ during the training includes several steps. The input $t$, taken from the parameter space $[t_{min},t_{max}]$ defines a laser sequence with $\Omega$ and $\delta$ fixed parameters followed by a measurement. The dynamics of the system is emulated and enables us to compute the probability distributions associated to this particular value of $t$ for the whole training part of the dataset. Finally, $F_1(t)$ is obtained by fitting the SVM with the kernel constructed from those probability distributions.}
	\label{fig:open-loop}
\end{figure*}

\section{Binary classification task}
\label{sec:Binary_class}

We now use the graph distance metric introduced in Eq.\,\eqref{eq:JS} to tackle a binary classification task on a dataset of chemical compounds called PTC-FM (Predictive Toxicity Challenge on Female Mice)\,\cite{Helma01,grakel}. The objective is to accurately predict the reactivity of chemical compounds (toxic/positive or non-toxic/negative) based on their structural properties. We will estimate the quality of the classification by using the F1 score $F_1=\frac{t_p}{t_p+\left(f_p+f_n\right)/2}$. Here, $t_p$, $f_p$ and $f_n$ are respectively the number of true positives, false positives and false negatives of the predicted distribution. Correctly predicting the toxicity of a compound (increasing $t_p$) leads to a better $F_1$ score. On the other hand, classifying a toxic compound as harmless (increasing $f_n$) or a harmless compound as toxic (increasing $f_p$) results in a lower score.    \\

\subsection{Quantum Evolution Kernel}
\label{sec:QEK_summary}
To realize the classification task on this dataset, we need to turn the quantum graph embedding introduced in Sec.\,\ref{sec:embeddings} into a kernel $K$. We follow the approach originally proposed in Ref.\,\cite{Henry21} to define this measure of similarity. We compute the distributions $\mathcal{P}$ of the total number of Rydberg excitations $\sum_j {\hat{n}}_j $ measured in the final state on graph $\mathcal{G}$ and we use again the Jensen-Shannon divergence from Eq.\,\eqref{eq:JS} to build a kernel out of those distributions.
\begin{gather}
	K(\Graph,\Graph') = \exp\left[- JS(\mathcal{P},\mathcal{P}')\right].
	\label{eq:kernel}
\end{gather}
This kernel is well-defined, {\it i.e.} the kernel matrix is always positive definite \cite{bai2013graph}. We feed this kernel to a Support Vector Machine (SVM) algorithm in order to discriminate between graphs of the dataset pertaining to one or the other class (see Appendix\,\,\ref{app:SVM}).

\begin{figure*}[t!]
	\centering
	\includegraphics[width=1\linewidth]{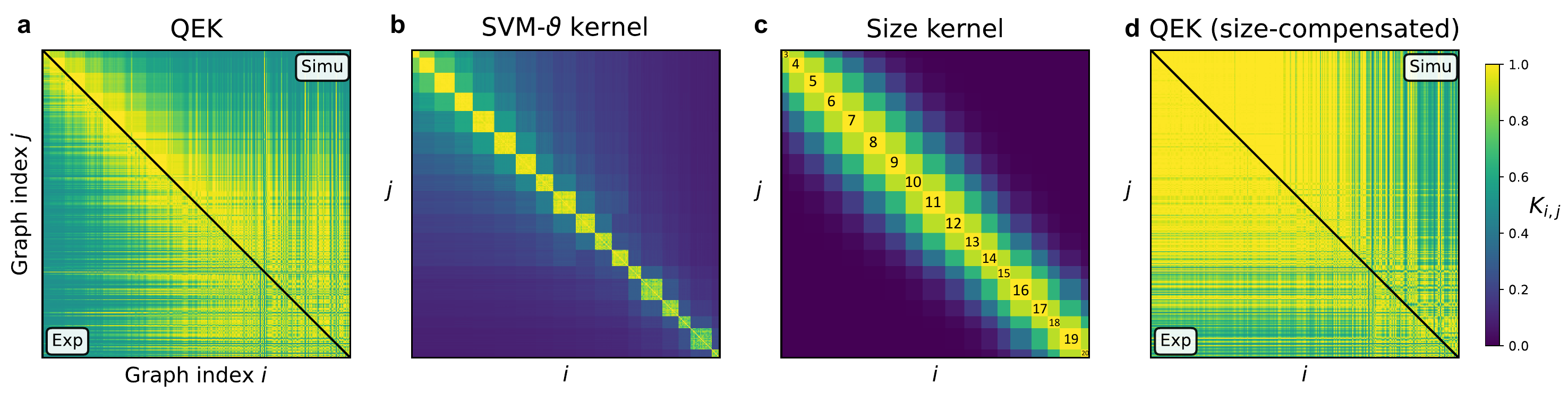}
	
	\caption{
	Each kernel is represented by a $M\times M$ matrix where $K_{i,j}=K(\Graph_i, \Graph_j)$ as defined in Eq.\,\eqref{eq:kernel}. The graph indices are sorted by increasing size. A separation (black line) is drawn between numerically simulated (top right) and experimentally measured (bottom left) QEK matrices.
	\textbf{a.} QEK kernel obtained using directly the raw distributions $\mathcal{P}_i$ and $\mathcal{P}_j$.
	\textbf{b.} Kernel obtained via SVM-$\vartheta$ method. 
	\textbf{c.} Size kernel obtained with $K^{\text{size}}(\Graph_i,\Graph_j)=\exp({-\gamma (|\Graph_i|-|\Graph_j|)^2})$ with $\gamma=0.1$.
	\textbf{d.} QEK kernel obtained using modified distributions $\mathcal{\tilde P}_i$ and $\mathcal{P}_j$, where graphs of smaller sizes are convoluted with binomial distributions when compared to larger graphs.}
	\label{fig:kernel_matrices}
\end{figure*}

\subsection{Dataset and mapping on hardware}
\label{sec:dataset}
In the original PTC-FM dataset, the $349$ molecules are represented under the form of graphs where each node is labeled by atomic type and each edge is labeled according to its bond type. We first truncate the dataset to small graph sizes in order to be able to train the kernel in reasonable time. For the $M=286$ remaining graphs of this dataset, we take into account the adjacency matrix of the graphs representing the compounds and discard the nodes and edges labels. 

Each node of a graph will be represented by a qubit in the QPU. We first need to determine the positions of these qubits.
To this end we design a local optimizer detailed in Appendix\,\ref{app:batching} to estimate in free space a preliminary 2D layout for each graph. Starting from a Reingold-Fruchterman layout\,\cite{Fruchterman91}, our optimizer minimizes the average distance between two connected nodes while maximizing the distances between unconnected nodes. This aims to implement an interaction term in Eq.\,\eqref{eq:Ham} that effectively reflects the graph topology. Then taking advantage of our ability to tailor the spatial disposition of the tweezers generated by a Spatial Light Modulator (SLM) to fit the optimized layout, we can replicate the graph in the hardware. Following a batching method also detailed in Appendix\,\ref{app:batching}, we group similar graphs and superimpose them on the same SLM pattern, effectively mapping the whole dataset on only $6$ different SLM patterns over a triangular grid. We therefore reduce the time needed to implement the whole dataset on the QPU.  

\subsection{Model training}
\label{sec:training}

To test the performance of our implementation, we perform a standard procedure called cross-validation. 
Cross-validation consists of dividing the dataset in $5$ equal parts called `splits', and using each split for testing while the rest of the dataset is used for training. During the training phase, we construct for each pulse duration $t$ the corresponding kernel and train a SVM model with it. We then evaluate the $F_1$-score on the part of the dataset that was left as a test set. We repeat the splitting $10$ times, and the cross-validation score is defined as the average of the $F_1$-score of each split ($50$ splits in total). We perform  a grid search on the penalty hyperparameter $C$ of the SVM on the range $[10^{-3}, 10^3]$ such that the final score of a given pulse is the best cross-validation score among the tested values of $C$, see Appendix \ref{app:SVM} for details.

Including graphs with sizes $|\Graph|\leq 20$, we numerically compute the score for a nearest-neighbor distance of $r_{NN} = 5.3\,\mu$m and a resonant constant pulse with fixed $\Omega/2\pi=1$ MHz, and we vary its duration between $t_{min}=0.1\,\mu$s and $t_{max}=2.5\,\mu$s. We select the optimal duration $T=0.66\,\mu s$ that exhibits the maximum $F_1$-score. We then implement this pulse on the QPU. The whole process is illustrated in Fig.\,\ref{fig:open-loop}.

\subsection{Classification results}
\label{sec:classification_results}

After a training of our model, we experimentally obtain an $F_1$-score of $60.4\pm 5.1\%$. For comparison purposes, we examine the performances of other kernels on this dataset: the Graphlet Sampling (GS), Random Walk (RW), Shortest Path (SP) and SVM-$\vartheta$ kernels, all these kernels being described in detail in Appendix \ref{app:classical_kernels}. The $F_1$-scores reached by the various kernels are collected in Table\,\,\ref{tab:score}. Obtained scores range from $49.8\pm 6.0\%$ up to $58.2\pm 5.5\%$. Those results show that the quantum kernel is competitive with standard classical kernels on this dataset. The SVM-$\vartheta$ kernel is found to be, among the classical kernels tested, the one with the best performance. As described in Appendix \ref{app:classical_kernels-svmtheta}, it is defined up to a choice of base kernel between real numbers, which gives it a certain degree of flexibility. \\

\begin{table}[t!]
\begin{tabular}{|c|c|}
\hline
Kernel          & $F_1$-score ($\%$)      \\ \hhline{|==|}
QEK         & $60.4 \pm 5.1$ \\ \hline
QEK (size-compensated)            & $45.1 \pm 3.7$ \\ \hline
SVM-$\vartheta$ & $58.2 \pm 5.5$ \\ \hline
Size            & $56.7 \pm 5.6$ \\ \hline
Graphlet Sampling              & $56.9 \pm 5.0$ \\ \hline
Random Walk              & $55.1 \pm 6.9$ \\ \hline
Shortest Path              & $49.8 \pm 6.0$ \\ \hline
\end{tabular}
\caption{$F_1$-score reached experimentally on the PTC-FM dataset by QEK ($\pm$ std. on the splits). In addition, the scores reached numerically by the classical kernels SVM$-\vartheta$, Size, Graphlet Sampling, Random Walk and Shortest-Path. The values reported are the average over a 5-fold cross-validation repeated 10 times.}
\label{tab:score}
\end{table}

\begin{figure}[t!]
    \centering
    \includegraphics[width=0.5\textwidth]{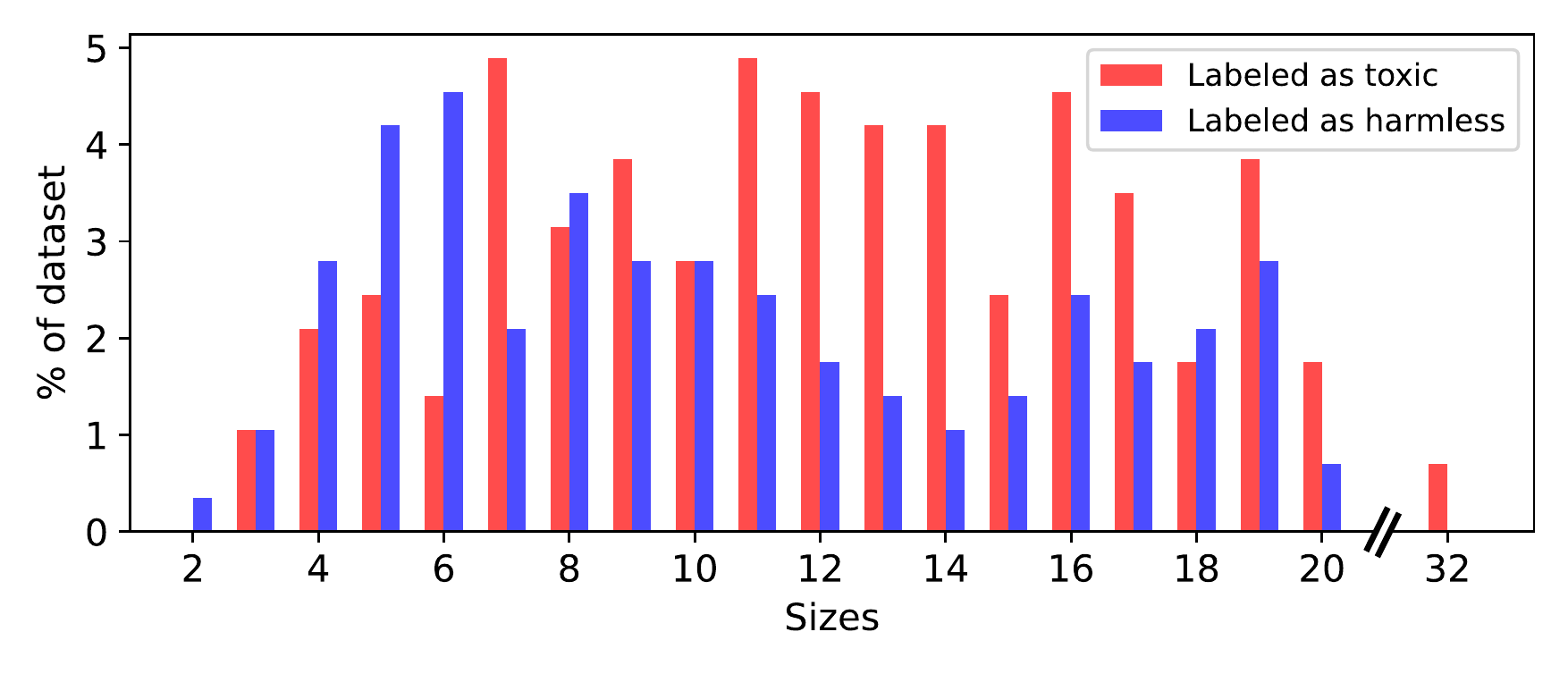}
    \caption{
    The PTC-FM dataset exhibits a strong size imbalance. For small number of nodes ($\lesssim$\,$10$) more graphs are labeled as harmless (blue) while it is the opposite for larger graphs, more prone to be labeled as toxic (red).}
    \label{fig:size_imbalance}
\end{figure}

We show in Fig.\,\,\ref{fig:kernel_matrices}\textbf{a} the kernel matrix associated with QEK, with indices sorted by increasing size of the graphs. Using the same noise model as in the previous section, we find adequate agreement between the numerically $\mathcal{P}^{num}$ and experimentally $\mathcal{P}^{exp}$ obtained data. Quantitatively, we make use of the JS divergence to estimate this agreement for any $\Graph_i$ and observe that $\langle JS(\mathcal{P}_i^{num},\mathcal{P}_i^{exp})\rangle_i\approx0.03\pm0.01 $ is one order of magnitude below $\langle JS(\mathcal{P}_i^{exp},\mathcal{P}_j^{exp})\rangle_{i\neq j}\approx0.33\pm0.01$. An interesting feature of both QEK and SVM-$\vartheta$ (Fig.\,\,\ref{fig:kernel_matrices}\textbf{b}) kernel matrices is the emergence of size-related diagonal blocks, signaling that the models identify the size of the graphs as an important feature for classification. Examining more closely the dataset, we indeed remark that the subset of PTC-FM that we used is significantly size imbalanced, as illustrated in Fig.\,\,\ref{fig:size_imbalance}. Since the graph size seems to be a relevant feature for this particular dataset, we define a Size kernel as a Gaussian in the size difference, which reaches an $F_1$-score of $56.7 \pm 5.6\%$. The corresponding kernel matrix is displayed in Fig.\,\,\ref{fig:kernel_matrices}\textbf{c} and exhibits a block-diagonal shape with a Gaussian tail. 

It is interesting to note that the quantum model was able to identify size as a relevant parameter for this dataset, leading to classification results which are on par with the best classical kernels.

Going forward, we modify the QEK procedure in order to make the kernel insensitive to size. To that end, we compare the measurement distributions obtained for different graph sizes using a convolution operation. Let us consider two graphs $\Graph_i$ and $\Graph_j$ of $N_i$ and $N_j=N_i+\Delta N>N_i$ nodes respectively; and note their respective observable distributions $\mathcal{P}_i$ and $\mathcal{P}_j$. From $\mathcal{P}_i$ we construct $\tilde{\mathcal{P}}_i = \mathcal{P}_i \star b_{\Delta N}^{(i/j)}$ the convolution of $\mathcal{P}_i$ and a binomial distribution :
\begin{equation}
    b_{\Delta N}^{(p)}(n) = \binom{\Delta N}{n} p^{n}(1-p)^{\Delta N-n}.
\end{equation}
$\tilde{\mathcal{P}}_i$ corresponds to the distribution one would get by adding to the graph $\Delta N$ non-interacting qubits, submitted to the same laser pulse as the other.
Each of these isolated qubits undergoes Rabi oscillations, induced by the applied pulse sequence. They are therefore measured either in $\ket{0}$ with probability $p$ or in $\ket{1}$ with probability $1-p$, where $p=\sin^2 (\pi \Omega T)$ ($\approx 0.768$ here).
We finally define the modified graph kernel as
\begin{gather}
	K_{conv}(\Graph_i,\Graph_j) = \exp\left[- JS(\tilde{\mathcal{P}}_i,\mathcal{P}_j)\right].
	\label{eq:convkernel}
\end{gather}
Using this procedure on the data obtained experimentally, we obtain the kernel matrix shown in Fig.\,\,\ref{fig:kernel_matrices}{\bf d}, with a corresponding $F_1$-score of $45.1 \pm 3.7\%$. If this size-compensated version of QEK had been implemented without interaction between atoms, its score would have been $42\%$, which is the lowest score reachable by any model. We therefore see that QEK cannot capture useful features beyond the graph size, meaning that we cannot use the interactions alone to produce an interesting kernel for the task at hand. While the size-compensated QEK does not give results that are comparable with classical kernels, we study in the following part its expressive power, and show that the geometry induced by this method is hardly reproducible by a classical kernel.

\subsection{Geometric test with respect to classical kernels}~
\label{sec:geo_diff}
In order to obtain an advantage over
classical approaches it is not sufficient to implement a quantum feature map based on quantum dynamics that are hard to simulate classically. As shown in~\cite{Huang2021}, classical ML algorithms can in certain instances learn efficiently from intractable quantum evolutions if they are allowed to be trained on data. The authors consequently propose another metric between kernels in the form of an asymmetric metric function called the geometric difference $g_{12}$. It compares two kernels $K_1$ and $K_2$ in the following way:
\begin{equation}
\label{eq:geom_diff_main_text}
    g_{12} = \sqrt{|| \sqrt{K_2} \left(K_1\right)^{-1} \sqrt{K_2} ||_\infty}
\end{equation}
where $||.||_{\infty}$ is the spectral norm. Intuitively, $g_{12}$ measures  the difference between how kernels $K_1$ and $K_2$ perceive the relation between data. Precisely, it characterizes the disparity regarding how each of them maps data points to their respective feature spaces. In our case, we take $K_1$ to be the size-compensated QEK $K_{conv}$, and $K_2$ is selected from a set of classical kernels. If the geometric difference is small, it means that there exists no underlying function mapping the data to the targets for which $K_{conv}$ outperforms the classical kernel. On the other hand, a high geometric difference between a quantum and a classical kernel guarantees that there exists such a function for which the quantum model outperforms the classical one. Estimating the geometric difference is therefore a sanity check to stating that the encoding of data to the feature space through the quantum kernel could not be closely replicated by a classical model.

We compute the geometric difference between QEK and various classical kernels over the PTC-FM dataset and report the results in Table\,\,\ref{tab:geodiff}. The threshold for a high geometric difference is typically taken to be $\sqrt{M}$, where $M$ is the size of the dataset. Here, the obtained $g_{12}$ is always far beyond $\sqrt{M}\sim10^1$, indicating that the embedding of data through our quantum-enhanced kernel is not trivial and cannot be replicated by a classical machine learning algorithm.

To summarize, while the $F_1$-score on PTC-FM is rather similar using quantum or classical models, we see nonetheless that the geometry created by our quantum model is non-trivial. A possible interpretation of the non-superiority of quantum approaches on PTC-FM would be that the relationship between the data and the targets is not better captured by our quantum model, although its feature space is not reproducible by classical means. To further confirm this understanding, we find a function that increases and even maximizes the utility of our rich quantum feature space. We build such a function by artificially relabeling the targets according to a procedure presented in \cite{Huang2021} and outlined in Appendix\,\ref{app:geo_diff}. We observe that QEK, without retraining, retains an $F_1$-score of around $99\%$ on the relabeled dataset, while the closest classical kernel reaches a score of at most $82\%$ even after retraining it on the new labels. The results are summarized in Table\,\,\ref{tab:gap}, where the difference in $F_1$-score between QEK and the various classical kernels is shown. 

\begin{table}[ht]
\begin{tabular}{|c|c|}
\hline
\multicolumn{2}{|c|}{Geometric Difference w.r.t. QEK} \\ \hhline {|==|}
\multicolumn{1}{|c|}{SVM-$\vartheta$}     & $10^3$    \\ \hline
\multicolumn{1}{|c|}{Size}                & $10^5$    \\ \hline
\multicolumn{1}{|c|}{Graphlet Sampling}                  & $10^4$    \\ \hline
\multicolumn{1}{|c|}{Random Walk}                  & $10^5$    \\ \hline
\multicolumn{1}{|c|}{Shortest Path}                  & $10^5$    \\ \hline
\end{tabular}
\caption{Order of magnitude of the geometric difference between QEK and various classical kernels.}
\label{tab:geodiff}
\end{table}

\begin{table}[ht]
\begin{tabular}{|c|c|}
\hline
\multicolumn{2}{|c|}{$F_1$-score gap (\%) w.r.t. QEK (relabeled)} \\ \hhline {|==|}
\multicolumn{1}{|c|}{SVM-$\vartheta$}     & $17.2\%$    \\ \hline
\multicolumn{1}{|c|}{Size}                & $17.8\%$    \\ \hline
\multicolumn{1}{|c|}{Graphlet Sampling}                  & $20.1\%$    \\ \hline
\multicolumn{1}{|c|}{Random Walk}                  & $17.3\%$    \\ \hline
\multicolumn{1}{|c|}{Shortest Path}                  & $18.2\%$    \\ \hline
\end{tabular}
\caption{Gap in $F_1$-score between QEK and various classical kernels after relabeling the dataset.}
\label{tab:gap}
\end{table}

In light of the geometric difference assessment and the observed gap of $F_1$-score between QEK and classical kernels on an artificial function, it remains an open question to generally characterize which types of dataset naturally offer a structure that better exploits the geometry offered by our quantum model, without requiring artificial tweaking of the labels. In the following section, we present a synthetic dataset on which QEK is able to outperform classical methods without any relabeling.

\subsection{Synthetic dataset}
\label{sec:synth_dataset}

\begin{figure}[h!]
	\centering
	\includegraphics[width=\linewidth]{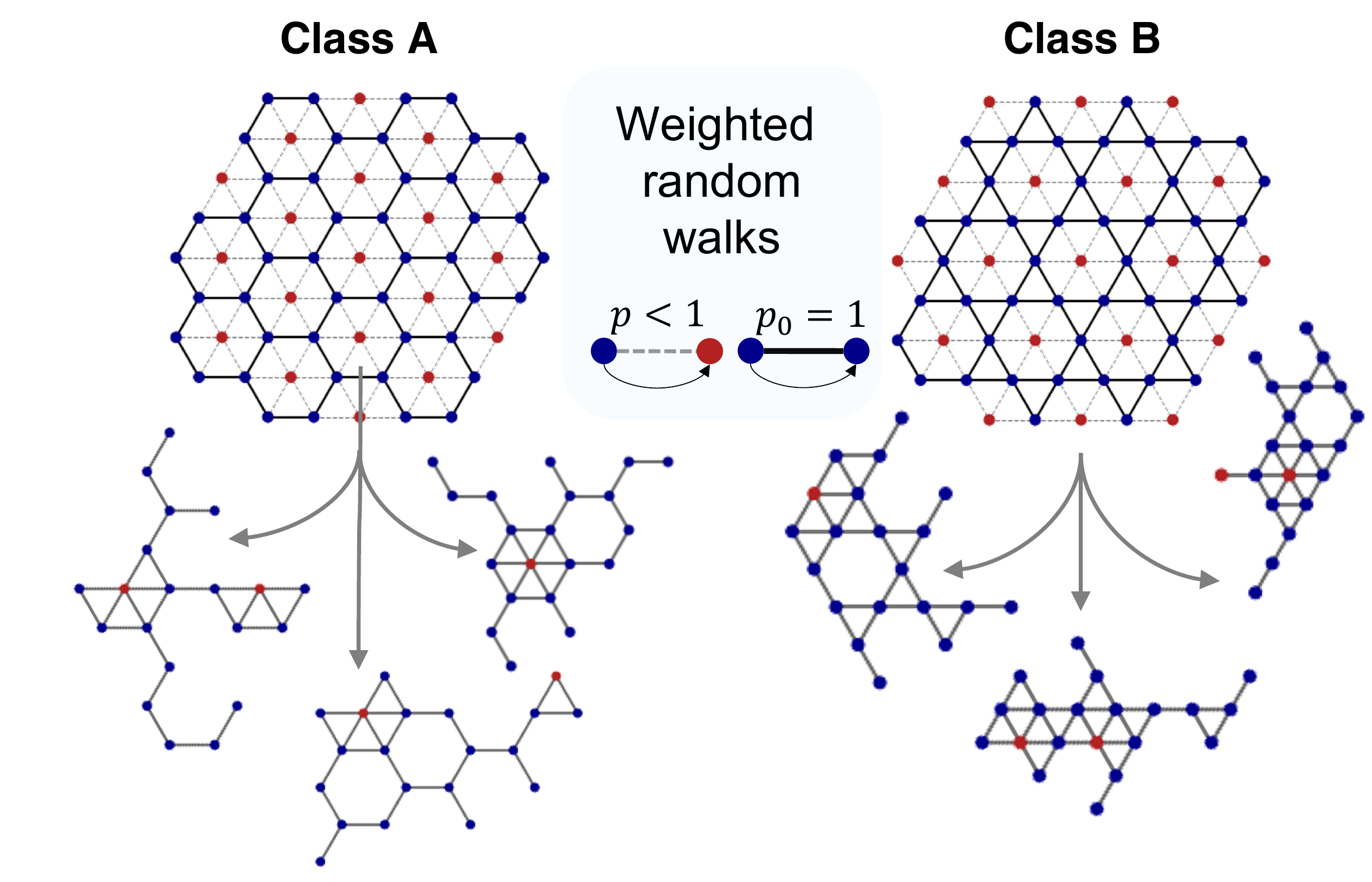}
	\caption{
	Graphs in Class A contains honeycomb sites (blue) with inclusions of non-honeycomb sites (red) with probability $p$. Graphs in Class B contains kagome sites (blue) with inclusions of non-kagome sites (red) with probability $p$. We show examples of generated graph with the aforementioned process.}
	\label{fig:synthetic}
\end{figure}

This binary classification dataset is created by sampling weighted random walks on a triangular lattice. In class A, sites belonging to a honeycomb-type sublattice are favored. They are explored with a weight $p_0=1$ while the rest of the triangular lattice sites are explored with a weight $p<1$. Class B is constructed in a similar fashion, but taking a kagome instead of a honeycomb sublattice. The construction of this artificial dataset is illustrated in Fig.\,\,\ref{fig:synthetic}.
In the case where $p=0$, the differences in their local structure make the two classes easily distinguishable.
However, with increasing $p$, their local structure becomes more and more similar, as additional triangular lattice sites are incorporated. When $p$ is large enough, a lot of triangular local substructures are shared by the two classes, rendering them potentially hard to distinguish by classical methods. At $p=1$, the underlying triangular lattice is explored uniformly, rendering the datasets indistinguishable.

Building on our ability to distinguish between graphs with similar local structure but globally distinct, we apply QEK on this synthetic dataset. We expect our method to be hardly affected by the presence of sparse defects and therefore be able to outperform classical approaches. 

\begin{figure}[h!]
	\centering
	\includegraphics[width=\linewidth]{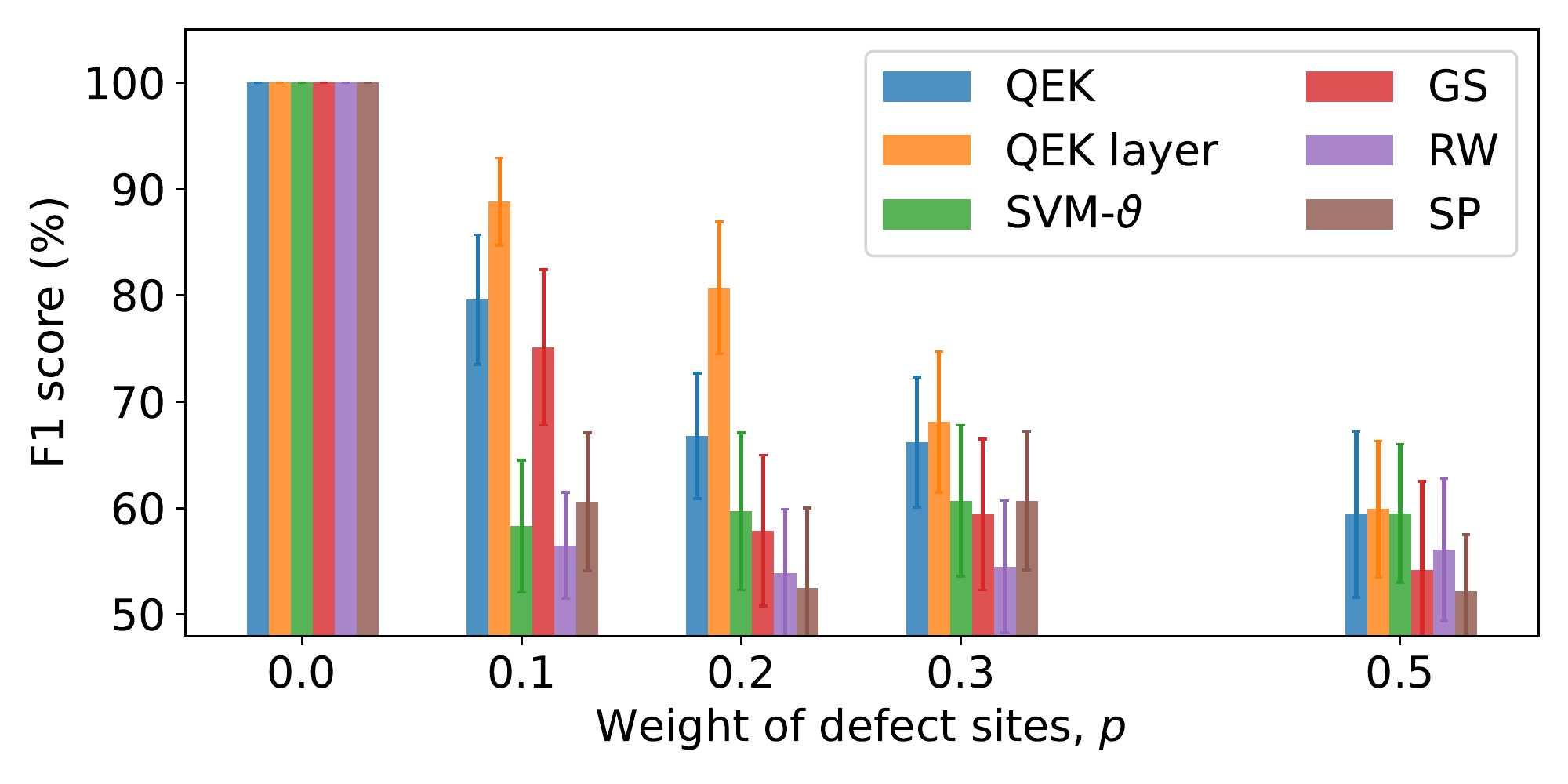}
	\caption{$F_1$-score (\%) reached on the synthetic dataset for different probabilities $p$ of including non-sublattice sites, by the Quantum Evolution Kernel (the alternate scheme is noted QEK layer) as well as by the best SVM-$\vartheta$,  GS, RW and SP kernels. The values reported are the average over a 5-fold cross-validation repeated 10 times. Each kernel reaches a $F_1$-score of 100\% when $p=0$.}
	\label{fig:synthetic_scores}
\end{figure}

We investigate numerically this assumption, for several values of $p$. In each case, we create $200$ graphs of $20$ nodes each, $100$ in each class. The graphs are mapped to a triangular lattice with $5$ µm spacing. Here, we consider two alternative schemes of pulse sequences. The first one remains almost the same as the experimentally implemented one, {\it i.e.} a unique resonant pulse of $\Omega/2\pi=2$  MHz with parameterized duration up to $8\,\mu$s. The second one is an alternate layer scheme with $4$ parameters as described in \cite{Henry21}, where we evaluate $500$ random values of the parameters and select the best one. The procedure is designed such that it would be directly implementable on the hardware, as we did for the PTC-FM dataset. We then compare the $F_1$-score reached by QEK to those reached by other classical kernels, namely: SVM-$\vartheta$, GS, RW and SP. The results are summarized in Fig.\,\,\ref{fig:synthetic_scores}. With decreasing proportion of defects, all methods perform increasingly better, as expected. Overall, regarding the mean $F_1$-score reached, the two QEK schemes outperform the four other classical kernels tested for all $p\leq0.5$. Noticeably, at $p=0.1$ (\textit{resp} $p=0.2$), the mean gap in $F_1$-score between the QEK scheme and the the best classical scheme is $4.5\%$ (\textit{resp} $7.1\%$) while the mean gap obtained with the alternate QEK scheme is even larger with $13.7\%$ (\textit{resp} $21\%$), thus showing that QEK can significantly surpass classical approaches on certain types of datasets. When adding too many defects, {\it i.e.} $p=0.5$, our quantum kernel exhibits similar performance to the SVM-$\vartheta$. 

\section*{Conclusion}
In this paper, we reported the implementation of a quantum feature map for graph-structured data on a neutral atom quantum processor with up to 32 qubits. We experimentally showed that this embedding was not only sensitive to local graph properties but was also able to probe more global structures such as cycles. This property offers a promising way to expand the capabilities of standard GNN architectures, which have been shown to have the same expressiveness as the Weisfeiler-Lehman (WL) Isomorphism test in terms of distinguishing non-isomorphic graphs \cite{Morris_Ritzert_Fey_Hamilton_Lenssen_Rattan_Grohe_2019,xu2018how}. For example, a Message Passing GNN will treat $\Graph_1$ and $\Graph_2$ shown in Fig.\,\,\ref{fig:res_WL_test}\textbf{a} in the same way, as they have the same local structure. Some properties of quantum-enhanced version of GNNs have been explored in \,\cite{Thabet2022}, by some of the authors of this paper.

We then used the quantum graph feature map for a toxicity screening procedure on a standard bio-chemistry dataset comprising 286 graphs of sizes ranging from 2 to 32 nodes. This procedure achieved a $F_1$-score of $60.4\pm5.1\%$, on par with the best classical kernels. We intentionally did not include GNNs in the benchmark, as they belong to another distinct family of models. Beyond this pure performance assessment, we showcased the potential advantage of using a quantum feature map through the computation of geometric differences with respect to said classical kernels, which are metrics evaluating the degree of similarity between the kernels' feature spaces. We showed that the quantum kernel captured features that are invisible to the classical kernels we considered. An artificial relabeling of the data enabled us to create a synthetic dataset for which the performances of the quantum kernel could not be matched. We also identified another dataset made of bi-partite 2D lattices, for which the quantum procedure exhibited superior performances.

This proof-of-concept illustrates the potential of quantum-enhanced methods for graph machine learning tasks. Our study paves the way for the incorporation of quantum-enhanced algorithms with standard ML solutions, aiming at constructing better tools for graph data analysis and prediction. Further work on more diverse datasets will be required to assess the viability of the approach compared to powerful state-of-the-art GNN architectures \citep{pmlr-v70-gilmer17a,ying2021transformers,rampavsek2022recipe,kreuzer2021rethinking}. Additionally, our results showcase the power and versatility of neutral atom QPUs, with their ability to change the register geometry from run to run. Going forward, the implementation of similar methods on non-local graphs could be envisaged by embedding them into three-dimensional registers \cite{3d_geometry} or moving the qubits throughout the course of the computation\,\cite{Bluvstein21moving}.

\section*{Acknowledgments}
We thank Jacob Bamberger, Lucas Beguin, Antoine Browaeys, Julia Cline, Luc Couturier, Romain Fouilland, Thierry Lahaye, Hsin-Yuan Huang, Christophe Jurczak, and Georges-Olivier Reymond for fruitful discussions.
\appendix

\section{Experimental setup}
\label{app:ES}

\begin{figure}[!b]
	\centering
	\includegraphics[width=0.5\textwidth]{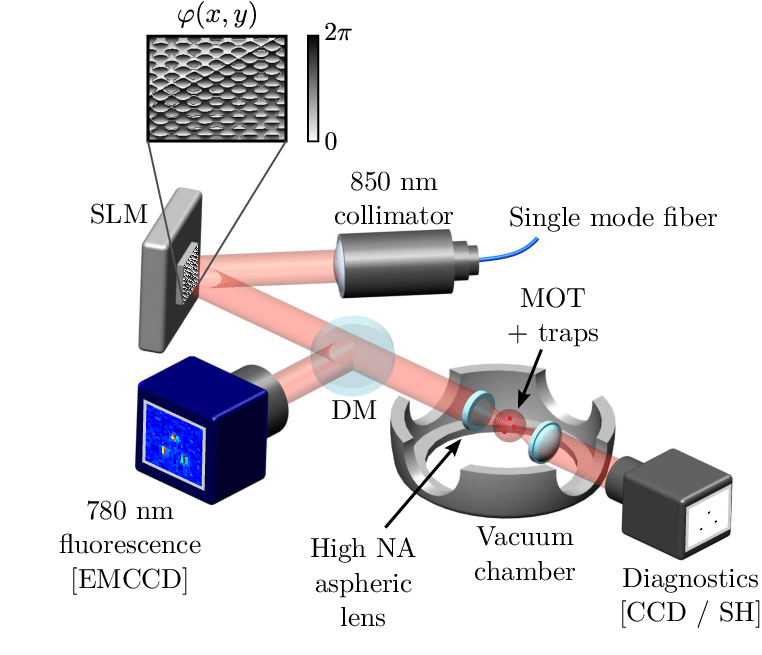}
	\caption{Figure from \cite{Nogrette14}. Microtraps for capturing single atoms are generated by using a SLM. A calculated phase pattern is printed on the $849$ $\mathrm{nm}$ laser beam and then focused by the first of two high numerical aperture lenses on the middle of a MOT. The atomic fluorescence at $780$ $\mathrm{nm}$ is reflected by a dichroic mirror (DM) and detected with an EMCCD camera. A second aspheric lens (identical to the first) collects the $849$ $\mathrm{nm}$ light for, three kind of images, layout loading (tweezers loading quality), register validation (rearrangement successfulness) and register readout (Rydberg excitation discrimination results). The transmitted beam is used for trap diagnostics via a CCD camera or a Shack-Hartmann wavefront sensor (SH).}
	\label{fig:setupiroise}
\end{figure}

The experimental setup is shown schematically in Fig.\,\ref{fig:setupiroise}. It features a magneto-optical trap (MOT), able to cool down and confine a cloud of $^{87}$Rb atoms in order to load an array of optical tweezers. They are created by shining a $849$ $\mathrm{nm}$ laser beam on a Spatial Light Modulator (SLM), and then focusing the beam to a small waist of $\sim 1$ $\mu$m with a high numerical aperture (NA) optical system inside the vacuum chamber. The loading of the optical tweezers is stochastic with a probability $\eta \approx 0.55$ of obtaining one atom per trap. Hence, at each repetition cycle of the experiment, we use a dynamical optical tweezer to move the atoms one by one in order to generate the targeted graph.

The atoms are embedded into a $10\,\mathrm{G}$ magnetic field that sets the quantization axis. The qubits are encoded into the ground state $\ket{0}=\ket{5S_{1/2},F=2,m_F=2}$ and a Rydberg state $\ket{1}=\ket{60S_{1/2}, m_J=1/2}$ of the atoms. They are initialized in the ground state by optical pumping. The qubit transition is then addressed by a two-photon laser excitation, via an intermediate state $6P_{3/2}$. The first (respectively second) photon excitation is generated by a $420\,\mathrm{nm}$ ($1013\,\mathrm{nm}$) $\sigma^+$-polarized ($\sigma^-$-polarized) laser beam with a $1/e^2$ waist radius of $260\,\mu$m ($180\,\mu$m). The two lasers being far-detuned from the intermediate state by $700\,\mathrm{MHz}$, we avoid spurious populating of this state and the three-level system can be faithfully approximated by an effective two-level system. The qubits state is readout in a single step by fluorescence imaging close to resonance at $780\,\mathrm{nm}$, using an EMCCD camera with an integration time of $20$ $\mathrm{ms}$.

A set of eight electrodes in an octupole configuration provides active control of the electric field environment around the Rydberg atoms. The durations and shapes of the Rydberg pulses are defined using acousto- and electro-optic modulators, in order to ensure the correct pulse-length used on the measurements.

\section{Noise model}
\label{app:noise_model}
Despite the precise calibration of the control devices which enable to monitor quantities such as the SLM pattern spacing or the pulse shapes, several experimental imperfections may alter the data measured on the experiment.  All experimental data obtained during this study, including those presented in Fig.\,\,\ref{fig:res_WL_test} and Fig.\,\,\ref{fig:kernel_matrices}, are uncorrected and thus needs to be benchmarked with respect to their simulated counterpart, taking into account the following main sources of noise.

First and foremost, due to the probabilistic nature of the quantum state and the limited budget of shots, measurements are subject to sampling noise. For instance, on average, each of the $25$ experimental points on Fig.\,\,\ref{fig:res_WL_test} is obtained using $600$ shots and the uncertainty related to this effect (vertical error bars) can be estimated using the Jackknife resampling method \cite{jackknife}. 

The finite sampling is also inherently flawed by several physical processes like atoms thermal motion, background-gas collisions or Rydberg state finite lifetime, whose effects can all be encompassed as first approximation into two detection error terms, $\varepsilon$ and $\varepsilon^\prime$. $\varepsilon$ (\textit{resp} $\varepsilon^\prime$) yield the probability to get false positive (\textit{resp} negative), {\it i.e.} measure an atom in $\ket{0}$ (\textit{resp} $\ket{1}$) as being in $\ket{0}$ (\textit{resp} $\ket{1}$). $\varepsilon$ can be measured with a regular release-and-recapture experiment and $\varepsilon^\prime$ with a more advanced method \cite{leseleucdekerouara:tel-02088297} involving $\pi$ and pushout pulses. To replicate the probabilistic effect of detection errors, the simulated distributions of bitstrings are altered using the following rule to compute the probability of measuring $j$ instead of $i$:
\begin{equation}
    P_{j|i}=\prod_k~(1-|i-j|_k)-(-1)^{|i-j|_k}\left[(1-i_k)\varepsilon+i_k \varepsilon^\prime\right].   
\end{equation}
$i,j\in\mathbb{B}^N$, $i_k=0$ (\textit{resp} $1$) if atom $k$ is in $\ket{0}$ (\textit{resp} $\ket{1}$). On our device, we measure $\varepsilon\approx3\%$ and $\varepsilon^\prime\approx8\%$; thus as an example, we can compute $P_{1001|0101}=\varepsilon\varepsilon^\prime(1-\varepsilon)(1-\varepsilon^\prime)\approx0.2\%$. Those detection errors can deeply modify the measured excitation distributions, with a noticeable effect shown on Fig.\,\,\ref{fig:res_WL_test}\textbf{b} at $t=0$ where the simulated $\langle n_j\rangle$ does not start at $0$ despite $\ket{\psi(t=0)}=\ket{0\hdots 0}$. 

Additional errors can also lead to decoherence in the system, affecting the atom dynamics in ways costly to emulate. For instance, since the Rydberg transition used is addressed by a two-photon process, misalignments and power fluctuations of the two lasers are twice as likely to occur. Atoms are subject to positional disorder between each shot and their finite velocities make them sensitive to the Doppler effect. Since taking all those effects into consideration becomes quickly intractable, they were individually simulated in order to assess their limited action on the implemented protocols. However, in order to replicate the experimental data presented in Fig.\,\,\ref{fig:res_WL_test}, we resort to an effective dephasing term of $2\pi\times 0.06$ MHz in the Master equation solver. This value was obtained by fitting damped Rabi oscillations measured on the same device. Thus, reaching similar behaviour within error bars between numerically simulated and experimentally obtained $JS(\mathcal{P}_1,\mathcal{P}_2)$ was achieved with no free parameter.\\ 

\section{Message passing neural networks}
\label{app:MPNN}
\begin{figure}[h!]
    \centering
    \includegraphics[width=.5\textwidth]{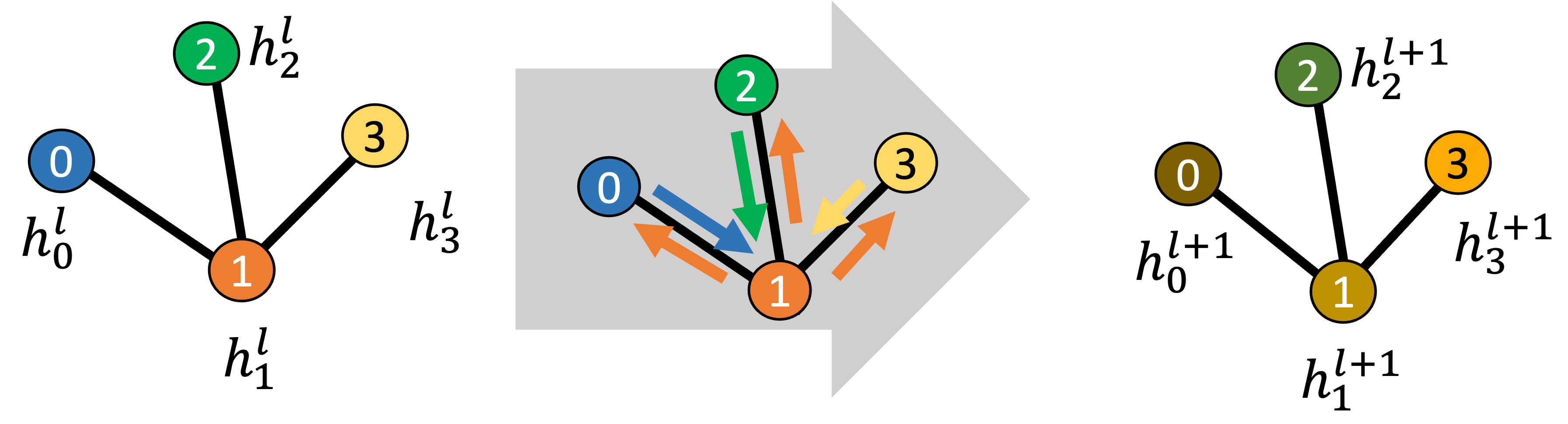}
    \caption{In message passing neural networks, node features vector $h_i^l$ are iteratively updated, from one layer $l$ to the next $l+1$, using only neighboring nodes, similarly to what is done in the WL test.}
    \label{fig:local_agg}
\end{figure}

Message passing neural networks (MPNN) \cite{pmlr-v70-gilmer17a} is a widely used family of graph neural networks. It was one of the first ones to be developed for graph-structured data, and is still one of the most successful \cite{MPNN_rev}. It consists of GNNs where the update is made only by aggregating the features of nearest neighbors.
In this scheme, the nodes features are multiplied by a trainable weight matrix at each layer, and each node aggregates as a 'message' the features of its neighbors, as illustrated in Fig.~\ref{fig:local_agg}. 

MPNN are closely related to Weisfeiler-Lehman algorithms. In particular, they have been proven to be at most as powerful in distinguishing graph structures~\cite{xu2018how}.
In their standard form, they are then also limited to capture only local features of graphs.

\section{Support Vector Machine (SVM) algorithm}
\label{app:SVM}

The SVM algorithm aims at splitting a dataset into two classes by finding the best hyperplane that separates the data points in the feature space, in which the coordinates of each data point (here each graph) is determined according to the kernel $K$.

For a training graph dataset $\left\{\Graph_i\right\}_{i=1\ldots M}$, and a set of labels ${\bf y}=\left\{y_i\right\}_{i=1\ldots M}$ 
(where $y_i=\pm1$ depending on which class the graph $\Graph_i$ belongs to), 
the dual formulation of the SVM problem consists in finding 
$\tilde{\boldsymbol{\alpha}}\in\mathcal{A}_C({\bf y})=\left\{\alphab\in[0,C]^{M}\right|\alphab^T{\bf y} = 0\}$
such that
\begin{gather}
    \frac{1}{2}\tilde\alphab^TQ\tilde\alphab - {\bf e}^T\tilde\alphab
    =\min_{\alphab\in\mathcal{A}_C({\bf y})}\;\left\{\frac{1}{2}\alphab^TQ\alphab - {\bf e}^T\alphab\right\}
\end{gather}
where ${\bf e}$ is the vector of all ones, $Q$ is a $M\times M$ matrix such that $Q_{ij} = y_i y_j K(\Graph_i,\Graph_j)$, and $C>0$ is the penalty hyperparameter, to be adjusted. Setting $C$ to a large value increases the range of possible values of $\alpha$ and therefore the flexibility of the model. But it also increases the training time and the risk of overfitting.

The data points for which $\tilde\alpha_i>0$ are called support vectors (SV). Once the $\alpha_i$ are trained, the class of a new
graph $\Graph$ is predicted by the decision function, given by:
\begin{align}
    \label{eq:SVM}
    y(\Graph) &= 
    \text{sgn}\left\{\braket{\phi(\Graph)|\phi_0}\right\}\\
    &=\text{sgn}\left\{\Sum_{i \in SV} y_i \tilde\alpha_i K(\Graph, \Graph_i)\right\}\label{eq:SVMb},
\end{align}
with 
\begin{equation}
    \label{eq:SVMphi0}
    \phi_0 = \Sum_{i \in SV} y_i \tilde\alpha_i\phi(\Graph_i).
\end{equation}
In this case, the training of the kernel amounts to finding the optimal feature vector $\phi_0$. It is worth noting that in many cases, Eq.\,\eqref{eq:SVMb} is evaluated directly, without explicitly computing $\phi_0$.

\section{Mapping and Batching}
\label{app:batching}

\begin{figure*}[ht]
    \centering
    \includegraphics[width=\textwidth]{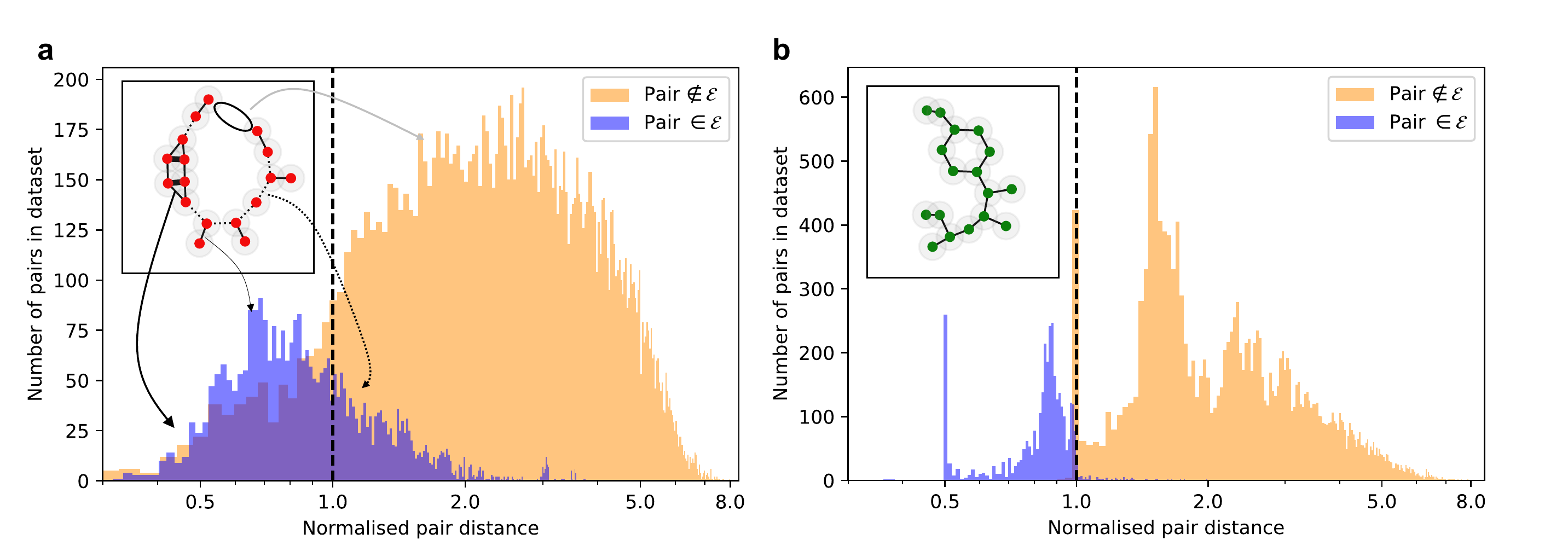}
    \caption{
    Histograms of normalised pairwise distances between atoms in the $286$ graphs of the truncated dataset when performing the embedding with \textbf{a.} only a Fruchterman-Reingold layout or \textbf{b.} when adding a local optimization step afterwards. For a given graph (insets), two atoms forming a pair $\in\mathcal{E}$ (blue) can be close enough to form a bond via interaction (plain) or too far, creating a missing bond (dotted). Likewise, two atoms forming a pair $\notin\mathcal{E}$ can be placed too close and form a fake edge (thick line).    }
    \label{fig:preprocess}
\end{figure*}

We present in detail our method to embed the graphs of the PTC-FM dataset. Let $\Graph=(\mathcal{V},\mathcal{E})$ be a graph of the dataset for which we have a layout of the nodes. Embedding the graph amounts to replace its nodes with atoms, the latter interacting between themselves with the $1/R^6$ dependence. Moving two atoms slightly apart can therefore drastically reduce their interaction strength but it remains non-zero. In order for the Hamiltonian to reflect the topology of $\Graph$, this $1/R^6$ dependence  needs to be approximated by the Heaviside function defined as:
\begin{equation}
    h(r) = \left\{
    \begin{array}{ll}
        \infty & \mbox{if } r\leq r_{b} \\
        0 & \mbox{else}
    \end{array}
\right.
\end{equation}

For the Heaviside approximation to be correct, we have to ensure that the largest distance between a pair sharing an edge in the graph is always far less than the shortest distance between a pair not sharing an edge. In other words, in theory, $\min \{U_{ij}, (i,j) \in \mathcal{E} \}/\max \{ U_{ij}, (i,j) \notin \mathcal{E} \}\gg1$. \\

We use a local optimizer to maximize this ratio and find good solutions in polynomial time. The method optimizes the position of each node in turn, depending on the previously mapped nodes and the presence of cycles in the graph. For the dataset used in this study, we achieve a significant increase of the mean ratio up to $16.8$, starting from $5.9$ with the classical Fruchterman-Reingold layout. We report that more than half the dataset exhibits a ratio higher than $10$ and less than $5\%$ of the dataset is embedded with some defects, {\it i.e.} a ratio smaller than $1$. We also assess the benefit of this approach in Fig.\,\,\ref{fig:preprocess} by comparing the distributions of distance of pairs $\in\mathcal{E}$ and pairs $\notin\mathcal{E}$ (\textbf{a}) before and (\textbf{b}) after the optimization. While some defects, such as fake or missing bonds, frequently appear in the pre-optimisation embedding, the optimised positions are constrained such that a clear cut is visible between the two distributions, easing the approximation. \\

In principle, we can program a different SLM pattern for the layout of each graph from the dataset. In practice however the SLM calibration step can be quite time-consuming, {\it i.e.} of the order of the minute. We can compare it to the duration of hundreds of shot, each of which consisting in applying a sequence and measuring a quantum state, performed at a frequency of $1$ Hz. Then for each graph, calibrating the SLM and obtaining the probability distribution take approximately the same order of time.\\

We therefore seek to regroup many graphs onto the same SLM pattern, to be able to reduce the number of calibrations needed for the whole dataset.
We do so by clustering the graphs according to similarities in their structures. Because the dataset consists in representations of organic molecules, many of the graphs share common structures. We thus focus on retrieving the presence and multiplicity of pentagons and hexagons. We then build a similarity measure between the graphs. For the pentagons for example, the similarity can be written under the form: 
\begin{equation}
    s(\Graph_1, \Graph_2) =1 - \exp(-\alpha |N^P_1 - N^P_2|)
\end{equation}
where $N^P$ represents the number of pentagons in $\Graph$ and $\alpha$ is a hyper-parameter. We then use a linear combination of similarity measures in order to build a similarity matrix between all graphs of the dataset. We then apply a k-means clustering algorithm \cite{k-means} using the similarity matrix in order to separate the graphs into different batches. Furthermore, since the laser power is distributed over all the traps, we want to reduce the total number of traps, in order to maximize the intensity provided to each trap. This ensures that the traps are deep enough to obtain a satisfying filling efficiency ($\sim55\%)$ over the whole pattern. For each batch, we thus apply the following mapping algorithm

\begin{algorithm}[H]
\caption{Creating a triangular SLM pattern by batching $M$ graphs}\label{alg:batching}
\begin{algorithmic}[1]
\Require Graphs $\{\Graph_1,\hdots, \Graph_M\}$ in sorted sizes and optimized positions $\{x_1,\dots, x_M\}$
\Ensure Single SLM pattern that embeds $M$ graphs with optimal positions on a triangular lattice.
\State  $ traps = \{\}$
\State for $i$ in range $1,\cdots,M:$ \\
        find $r_{\Graph_i}=\{r_1, \dots, r_{|\Graph_i|}\}$ triangular grid points that best conserve the pairwise distances between points in $x_i$ and maximizes overlap with existing $traps$. \\
        $traps \leftarrow traps + r_{\Graph_i}$ \textbackslash $traps$
\State if $|traps| < 2|\Graph_M|$, add additional random triangular grid points to guarantee the filling property for re-arrangement.
\end{algorithmic}
\end{algorithm}

\begin{figure}[h!]
	\centering
	\includegraphics[width=0.9\linewidth]{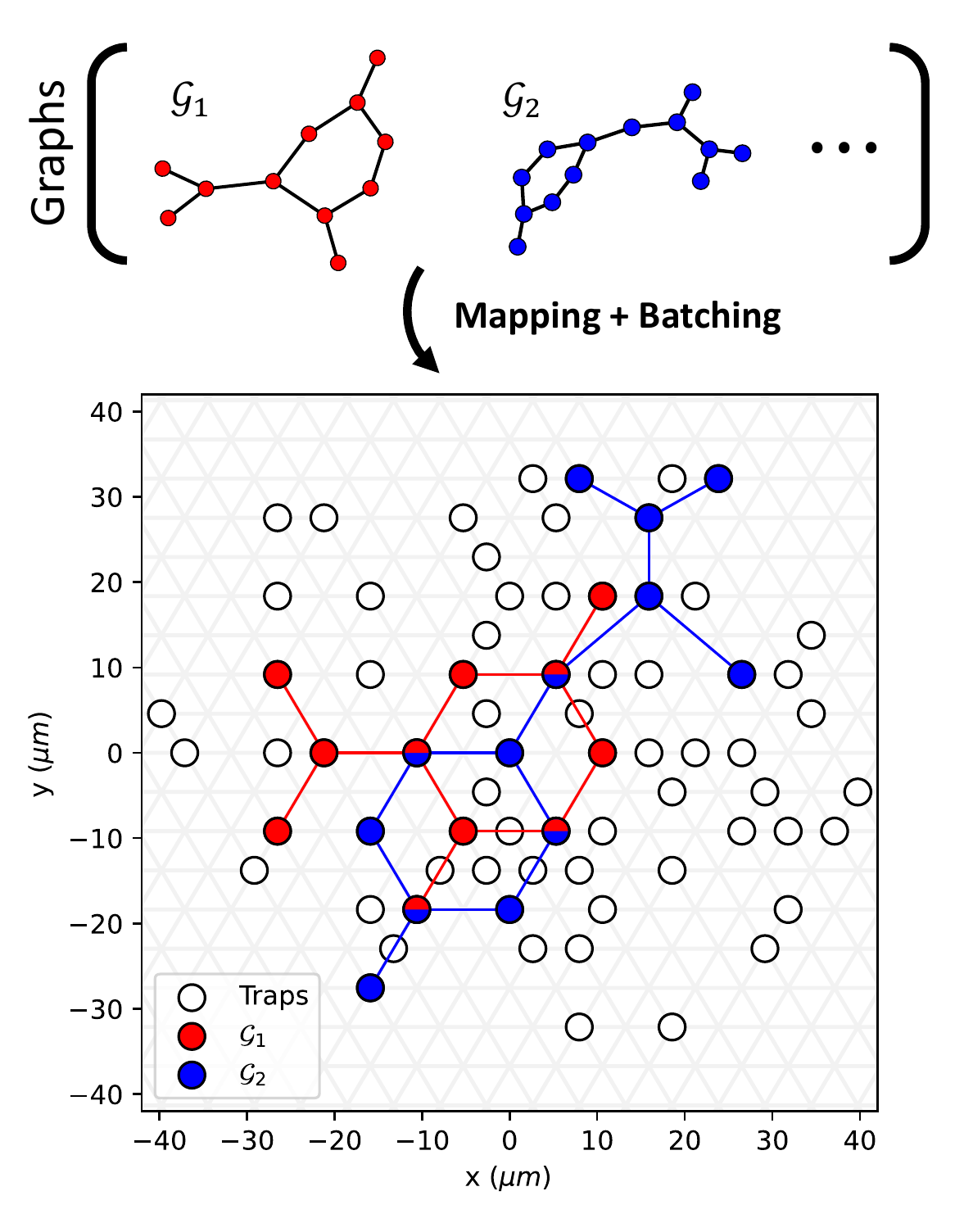}
	\caption{
	A family of 66 graphs, ranging in sizes from $4$ to $19$ nodes, is mapped and batched to the same SLM pattern (white dots) over a triangular grid with spacing $5.3\,\mu$m. The traps used when implementing $\mathcal{G}_1$ ($\mathcal{G}_2$) are colored in red (blue). The bi-colored traps are those used for both graphs.}
	\label{fig:mapping}
\end{figure}

We successfully map the entire dataset of $286$ graphs into only $6$ SLM patterns. For example, we batch $66$ graphs together onto the $71$-trap SLM pattern presented in Fig.\,\,\ref{fig:mapping}. On average, the $6$ SLM patterns use $70$ traps each to encode $48$ graphs each.

\section{Classical graph kernels}
\label{app:classical_kernels}

A variety of classical kernels that do not require node or edge attributes are used in the main text to compare the performance of QEK on the PTC-FM dataset. In the following, a brief description of each is given.

\subsection{SVM-$\vartheta$ kernel}
\label{app:classical_kernels-svmtheta}

The SVM-$\vartheta$ kernel was proposed as an alternative to the more computationally intensive Lovasz-$\vartheta$ kernel. Both $\vartheta$ kernels leverage the so-called orthogonal representation of a graph. Given a graph $\Graph=(\vertices,\edges)$, the orthogonal representation is an assignment of unit vectors $\{\mathbf{u}_i\}$ to each node of the graph, subject to the constraint that unit vectors associated to vertices that are not joined by an edge are orthogonal: $\langle \mathbf{u}_i, \mathbf{u}_j\rangle = 0$ if $\{i,j\} \notin \edges$.

Orthogonal representations are not unique, but there is a particular representation associated with the $\vartheta$ number \cite{lovasz} of a graph. Given a graph $\Graph=(\vertices,\edges)$ with $n$ vertices, denote $U_\Graph$ an orthogonal representation of $\Graph$, and $C$ the space of unit vectors in $\mathbb{R}^n$. The $\vartheta$ number is defined as:
\begin{equation}\label{eq:theta}
    \vartheta(\Graph) := \min_{\mathbf{c} \in C} \min_{U_\Graph} \max_{\mathbf{u}_i \in U_\Graph} \frac{1}{\langle \mathbf{c},  \mathbf{u}_i \rangle^2}.
\end{equation}
From now on, we will always be referring to the particular orthogonal representation $U_\Graph$ that minimizes (\ref{eq:theta}).

Now consider a subset of vertices $B\subset \vertices$, and call $U_{\Graph|B}$ the orthogonal representation obtained from $U_\Graph$ by removing the vectors that are not in $B$:
\begin{equation}
    U_{\Graph|B} := \{ \mathbf{u}_i \in U_\Graph: \ i\in B\}.
\end{equation}
Note that $U_{\Graph|B}$ preserves the global properties encoded in $U_\Graph$ through the orthogonal constraint, and that $U_{\Graph|B}$ is not in general the orthogonal representation of the subgraph of $\Graph$ containing only the vertices in $B$. Define the $\vartheta_B$ number:
\begin{equation}\label{eq:theta_B}
    \vartheta_B(\Graph) := \min_{\mathbf{c} \in C}  \ \max_{\mathbf{u}_i \in U_{\Graph|B}} \ \frac{1}{\langle \mathbf{c},  \mathbf{u}_i \rangle^2}.
\end{equation}
We are ready now to give the definition of the Lovasz-$\vartheta$ kernel. Given two graphs $\Graph_1=(\vertices_1,\edges_1)$, $\Graph_2=(\vertices_2, \edges_2)$, define:
\begin{equation}\label{eq:ltkernel}
    K_\text{Lo}(\Graph_1, \Graph_2) := \sum_{B_1 \subset V_1}\sum_{B_2 \subset V_2} \delta_{|B_1|, |B_2|}\ \frac{1}{Z} \ k\left(\vartheta_{B_1}, \vartheta_{B_2}\right)
\end{equation}
where $Z = \binom{|V_1|}{|B_1|}\binom{|V_2|}{|B_2|}$, $\delta$ is the Kronecker delta, and $k$ is a freely specifiable kernel (called base kernel) from $\mathbb{R} \times \mathbb{R} $ to $\mathbb{R}$.

The SVM-$\vartheta$ kernel is defined as (\ref{eq:ltkernel}), but it uses an approximation for the $\vartheta$ numbers. Consider a graph $\Graph$ with $n$ vertices and adjacency matrix $A$, and let $\rho \ge -\lambda$, where $\lambda$ is the minimum eigenvalue of $A$. The matrix
\begin{equation}
    \kappa := \frac{1}{\rho} A + I
\end{equation}
is positive semi-definite. Define the maximization problem:
\begin{equation}\label{eq:maxsvm}
    \max_{\alpha_i \ge 0} \ 2 \sum_{i=1}^n \alpha_i - \sum_{i,j=1}^n \alpha_i \alpha_j \kappa_{ij}.
\end{equation}
If $\{\alpha^*_i\}$ are the maximizers of (\ref{eq:maxsvm}), then it can be proven that on certain families of graphs the quantity $\sum_i \alpha^*_i$ is with high probability a constant factor approximation to $\vartheta(\Graph)$:
\begin{equation}
    \vartheta(\Graph) \le \sum_{i=1}^n \alpha^*_i \le \gamma \vartheta(\Graph)
\end{equation}
for some $\gamma$. The SVM-$\vartheta$ kernel then replaces the $\vartheta_B$ numbers on subgraphs with:
\begin{equation}
    \vartheta_B(\Graph) \rightarrow \sum_{j \in B} \alpha^*_j.
\end{equation}
The SVM-$\vartheta$ kernel requires a choice of base kernel $k: \mathbb{R} \times \mathbb{R} \rightarrow \mathbb{R}$. We choose a translation invariant universal kernel \cite{JMLR:v7:micchelli06a} $k(x,y) = (\beta + ||x-y||^2)^{-\alpha}$, where $\alpha$ and $\beta$ are two trainable hyperparameters.

\subsection{Size kernel}
Given two graphs $\Graph_1 = (\vertices_1, \edges_1)$ and $\Graph_2 = (\vertices_2, \edges_2)$, the Size kernel is defined as:
\begin{equation}
    K_\text{size}(\Graph_1,\Graph_2) := e^{-\gamma \left( |\vertices_1| - |\vertices_2|  \right)^2}     
\end{equation}
with a choice of hyperparameter $\gamma>0$.

\subsection{Graphlet Sampling kernel}
Let $\Graph=(\vertices,\edges)$ and $H=(\vertices_H, \edges_H)$ be two graphs. We say that $\mathcal{H}$ is a subgraph of $\Graph$ if there exists an injective map $\alpha: \vertices_\mathcal{H} \rightarrow \vertices$ such that $(u,v)\in \edges_H \iff (\alpha(u), \alpha(v)) \in \edges$. In general it might be possible to map $\mathcal{H}$ into $\Graph$ in several different ways, {\it i.e.} the mapping $\alpha$, if it exists, is not necessarily unique.

Given two graphs $\Graph_1=(\vertices_1, \edges_1)$ and $\Graph_2=(\vertices_2, \edges_2)$, the idea behind the Graphlet kernel is to pick an integer $k < \min \{|\vertices_1|, |\vertices_2|\}$, enumerate all possible graphs of size $k$ and find the number of ways they can be mapped to $\Graph_1$ and $\Graph_2$. Denote by $f^{(k)}_{\Graph_i}$ the vector where each entry counts the way a specific graph of size $k$ can be mapped as a subgraph of $\Graph_i$. A kernel can then be defined as the dot product $f^{(k)}_{\Graph_1} \cdot f^{(k)}_{\Graph_2}$ between the two vectors.

The complexity of computing such a kernel scales as $O(n^k)$, as there are $\binom{n}{k}$ size-$k$ subgraphs in a graph of size $n$. For this reason it is preferable to resort to sampling rather than complete enumeration~\cite{graphletsampling}. Given a choice of integer $N$, graphs $g_1, \ldots, g_N$ of size between 3 and $k$ are randomly sampled. The number of ways each $g_i$ can be mapped as a subgraph of $\Graph_j$ is computed and stored in a vector $f_{\Graph_j}$, and the Graphlet Sampling kernel is defined as the dot product:
\begin{equation}
    K_\text{GS} (\Graph_1, \Graph_2) := f_{\Graph_1} \cdot f_{\Graph_2}
\end{equation}
To account for the different size of $\Graph_1$ and $\Graph_2$, each vector can be normalized by the total number of its subgraphs.

\subsection{Random Walk kernel}

The Random Walk kernel is one of the oldest and most studied graph kernels~\cite{randomwalk}. Given two graphs $\Graph_1=(\vertices_1, \edges_1)$ and $\Graph_2=(\vertices_2, \edges_2)$, the idea is to measure the probability of simultaneous random walks of a certain length between two vertices in $\Graph_1$ and $\Graph_2$.

Simultaneous random walks can be conveniently encoded in powers of the adjacency matrix on the product graph. The product graph $\Graph_1 \times \Graph_2 = \Graph_\times = (\vertices_\times, \edges_\times)$ is defined as follows:
\begin{align}
    \vertices_\times &:= \{(u_i, u_r) \mid u_i \in \vertices_1, \ u_r \in \vertices_2\} \\
    \edges_\times &:= \{ \big((u_i, u_r), (v_j, v_s)\big) \mid (u_i, v_j) \in \edges_1, \notag \\
    &\quad \quad \quad \quad \quad \quad \quad \quad (u_r, v_s) \in \edges_2\}.
\end{align}
In other words, an edge in the product graph indicates that an edge exists between the endpoints in both $\Graph_1$ and $\Graph_2$. If $A_\times$ is the adjacency matrix of the product graph, then the entries of $A_\times^k$ indicate the probability of a simultaneous random walk of length $k$ between two vertices $u_i, v_j \in \vertices_1$ and $u_r, v_s \in \vertices_2$.

If $p, q \in \mathbb{R}^{|\vertices_\times|}$ are vectors representing the probability distribution of respectively starting or stopping the walk at a certain node of $\vertices_\times$, the first idea for a kernel would be to compute the sum $\sum_k q^\text{T}  A_\times^k p $, which however may fail to converge. A simple modification to make the sum convergent is to choose an appropriate length-dependent weight $\mu(k)$:
\begin{equation}
    K(\Graph_1, \Graph_2) := \sum_{k=0}^\infty \mu(k) \ q^\text{T} A_\times^k p.
\end{equation}
The Geometric Random Walk kernel is obtained by choosing the weights to be the coefficients of a geometric series $\mu(k) = \lambda^k$, and $p, q$ to be uniform. If $\lambda$ is tuned in such a way as to make the series convergent, the kernel reads:
\begin{equation}
    K_\text{RW}(\Graph_1, \Graph_2) := \sum_{k=0}^\infty \lambda^k \ e^\text{T} A_\times^k e = e^\text{T} \left(I - \lambda A_\times \right)^{-1} e
\end{equation}
where $e$ denote vectors with all the entries equal to 1.

The cost of matrix inversion scales as the cube of the matrix size. If $|\vertices_1| = |\vertices_2| = n$, then the cost of the algorithm scales as $O(n^6)$, as it involves the inversion of an adjacency matrix of size $n^2 \times n^2$. Several methods are proposed in~\cite{rwefficient} to make the computation faster. The Spectral Decomposition method in particular allows to reduce the complexity for unlabeled graphs to $O(n^3)$. Essentially, one exploits the fact that the adjacency matrix of the product graph can be decomposed in the tensor product of the individual adjacency matrices:
\begin{equation}
    A_\times = A_1 \otimes A_2
\end{equation}
which allows to diagonalize each $n\times n$ adjacency matrix in $O(n^3)$ time and perform the inversion only on the diagonal components.

\subsection{Shortest Path kernel}
Given a graph $\Graph=(\vertices,\edges)$, an edge path between two vertices $u, v \in \vertices$ is a sequence of edges $(e_1, \ldots, e_n)$ such that $u\in e_1$, $v\in e_n$, $e_i$ and $e_{i+1}$ are contiguous ({\it i.e.} they have one of the endpoints in common) and $e_i \neq e_j$ for $i\neq j$. Computing the shortest edge path between any two nodes of a graph can be done in polynomial time with the Dijkstra~\cite{dijkstra} or Floyd-Warshall~\cite{flowar} algorithms, which makes it a viable feature to be probed by a graph kernel.

The first step of the Shortest Path kernel is to transform the graphs into shortest path graphs. Given a graph $\Graph=(\vertices,\edges)$, the shortest path graph $\Graph^S = (\vertices^S, \edges^S)$ associated to $\Graph$ is defined as:
\begin{align}
    \vertices^S &= \vertices \\
    \edges^S &= \{ (u, v) \mid \exists \text{ an edge path } (e_1, \ldots, e_n) \notag \\
    &\text{ between $u$ and $v$ in $\Graph$} \}
\end{align}
In addition, to each edge $e\in \edges^S$ a label $l(e)$ is assigned given by the length of the shortest path in $\Graph$ between its endpoints. The Shortest Path kernel is then defined as:
\begin{equation}
    K_\text{SP}(\Graph_1, \Graph_2) := \sum_{e \in \edges^S_1} \sum_{p \in \edges^S_2} k(e, p)
\end{equation}
with $k$ being a kernel between edge paths such as the Brownian bridge kernel:
\begin{equation}
    k(e,p) := \max\{0, \ c - |l(e)-l(p)|\}
\end{equation}
for a choice of $c$.

\section{Geometric difference and maximum quantum-classical separation}
\label{app:geo_diff}

Given two kernel functions $K_1$ and $K_2$, the geometric difference $g(K_1 || K_2) = g_{12}$ described in \cite{Huang2021} is an asymmetric distance function that quantifies whether or not the kernel $K_2$ has the potential to resolve data better than $K_1$ on some dataset. In its simplest form, the geometric difference is defined as:
\begin{equation}\label{eq:geom_diff_simple}
    g_{12} = \sqrt{|| \sqrt{K_2} \left(K_1\right)^{-1} \sqrt{K_2} ||_\infty}
\end{equation}
where $||\cdot||_\infty$ denotes the spectral norm.

The geometric difference becomes an especially useful metric when $K_1=K_C$ is a classical kernel and $K_2=K_Q$ is a quantum kernel. If $N$ is the size of the dataset, a value of $g_{CQ}$ of order $\sqrt{N}$ or greater indicates that the geometry of the feature space induced by the quantum kernel is rich enough to be hard to learn classically, and the quantum kernel can potentially perform better than classical kernels. In that case, it is possible to artificially relabel the dataset in order to maximally separate the kernels' performance. Such a relabeling process is a constructive proof of the existence of a certain dataset on which one kernel performs much better than the other. If $v$ is the eigenvector of $\sqrt{K_2} \left(K_1\right)^{-1} \sqrt{K_2}$ corresponding to the eigenvalue $g_{12}^2$, the vector of new labels is given by $ y_\text{new} = \sqrt{K_2} v$.

When dealing with a finite amount of training data, equation (\ref{eq:geom_diff_simple}) should be regularized in order to stabilize the inversion of $K_1$. The regularized expression reads:
\begin{equation}\label{eq:geom_diff_reg}
    g_{12}(\lambda) = \sqrt{|| \sqrt{K_2} \sqrt{K_1} \left(K_1 + \lambda I\right)^{-2} \sqrt{K_1} \sqrt{K_2} ||_\infty}
\end{equation}
where $\lambda$ is the regularization parameter. The geometric difference $g_{12}(\lambda)$ has a plateau for small $\lambda$, when the regularization parameter becomes smaller than the smallest eigenvalue of $K_1$, and decreases for increasing $\lambda$. The effect of $\lambda$ is to introduce a certain amount of training error. The training error can be upper bounded by a quantity proportional to:
\begin{equation}\label{eq:tra_err_reg}
    g_\text{tra}(\lambda)^2 = \lambda^2 || \sqrt{K_2} \left(K_1 + \lambda I\right)^{-2} \sqrt{K_2} ||_\infty.
\end{equation}
Practically, one should look at the regime where $g_{12}$ has not plateaued but the training error is still small enough.

A regularization should be introduced also in the relabeling procedure. The new labels are taken to be $ y_\text{new} = \sqrt{K_Q} v$, where $v$ is the eigenvector of the regularized matrix $$\sqrt{K_Q} \sqrt{K_C} \left(K_C + \lambda I\right)^{-2} \sqrt{K_C} \sqrt{K_Q}$$ corresponding to the eigenvalue $g_{12}(\lambda)^2$.

\bibliographystyle{IEEEtran}
\bibliography{refs}

\end{document}